\documentclass{aa}  
\pdfoutput=1
\usepackage{graphicx}
\usepackage{txfonts}
\usepackage{hyperref}
 \usepackage{amsmath}
 \usepackage{natbib}
 \bibpunct{(}{)}{;}{a}{}{,} 
 \usepackage{epsfig}
 \usepackage{color}
 \usepackage{lscape}  
 \graphicspath{{figs/}}
\usepackage{longtable}
\usepackage[normalem]{ulem}
\usepackage{amstext}

\begin{document}

\title{Hidden power of near-infrared data for the study of young
  clusters: Illustrative case of RCW 38}
\titlerunning{Hidden power of near-infrared data}

   \author{Joana Ascenso\inst{1,2}}

   \institute{Departamento de Engenharia F\'isica, Faculdade de
     Engenharia, Universidade do Porto, Rua Dr. Roberto Frias,
     4200-465 Porto, Portugal \and CENTRA, Instituto Superior Tecnico,
     Universidade de Lisboa, Av. Rovisco Pais 1, 1049-001, Lisbon,
     Portugal }

   \date{Accepted for publication in Astronomy \& Astrophysics.}

\newcommand{\ccd}{$(J-H)$ {\it \text{\text{\textup{vs}}}} $(H-K_s)$ color-color diagram}
\newcommand{\hk}{$(H-K_s)$}
\newcommand{\jh}{$(J-H)$}
\newcommand{\vs}{{\it vs}}

\abstract {Studies of star formation rely heavily on observations in
  the near-infrared, but they typically need information from other
  wavelengths for interpretation. We show that we can infer distances
  and estimate the membership of young stellar objects for young
  clusters independently using (ground-based) near-infrared, $J$, $H,$
  and $K_S$ broadband data alone. We also show that we can estimate a
  lower limit for the fraction of sources with $2.2~\mu$m excess
  emission with a sensitivity comparable to that of mid-infrared space
  data, but with better resolution and fewer biases. Finally, we show
  that the typical methods for inferring masses from these data may
  produce substantially unreliable results. This method is applied to
  the young, massive cluster RCW 38, for which we estimate a distance
  of 1.5 kpc and a $K_S$-band excess fraction larger than 60\%.}


   \maketitle
%

\section{Introduction}\label{sec:introduction}

Understanding the process of star formation from molecular clouds
depends critically on our ability to determine physical parameters for
the young stellar population. From the calibration of stellar
evolution models to the reconstruction of the star formation history
of a given region, or to the determination of the star formation
efficiency or the formation of planetary systems, understanding the
young populations is paramount for constraining the processes that
transform gas into stars. The new generation of observing facilities,
in particular the giant telescopes, will allow detailed studies of
star-forming regions up to much larger distances than what is possible
today, namely through near-infrared instrumentation, increasing our
sample of objects exponentially and the range of parameters they
cover, and making it increasingly important to have a reliable handle
on the parameters of young stellar objects (YSOs).

Near-infrared (NIR) observations have gained a large predominance in
the study of star-forming regions and young clusters over the past
three decades because they take advantage of the relative
transparency of the atmosphere in these wavelengths and because
current instrumentation can reach very high resolutions and
sensitivities from the ground. Coupled with the lower extinction
from molecular clouds in the NIR when compared to the optical
\citep[e.g.,][]{Rieke85}, this makes these observations ideal for the study
of young populations that are still embedded in molecular gas and
dust. These observations -- in particular broadband photometry --
are considered insufficient for characterizing these populations, however, because
they need to rely on other wavelengths to determine important
parameters, such as distances and membership of individual sources,
without which the analysis is significantly hampered.

In this paper, we show that NIR photometric data can be used
independently to estimate these two key parameters -- distance and
membership -- for young clusters, opening new possibilities for the
study of star formation. We also show how these data can be used to
restrict relevant physical parameters, namely the cluster excess
fractions, and the excess and mass of YSOs.

The mass of a YSO is one of its most meaningful physical properties
because it allows both understanding individual objects and the study
of stellar populations. Currently, the most accurate way to estimate
masses of YSOs is through spectroscopy of individual objects, but this
approach is inefficient and observationally expensive, especially for
large populations and for regions with strong contamination from
unrelated objects. For this reason, multiwavelength photometric
surveys are widely used instead to assess the masses of YSO
populations \citep[e.g.,][]{Hillenbrand00,Muench02, Stolte06,
  Ascenso07, Ascenso07b, Harayama08, Preibisch11, Rochau11, Habibi13,
  Scholz13, Neichel15, Drass16, Muzic17}. They allow characterizing a
large number of objects simultaneously, and they need shorter
observation times to reach equally faint objects. Photometry carries
less information than spectroscopy, but it is the only viable option
for reasonably complete studies of populous star-forming regions or
regions that are too distant.

The fraction of sources with excess from circumstellar material in a
young stellar cluster is another very telling property of its
evolution. Circumstellar material is observed around YSOs, mostly in
the form of disks by the time these objects are visible in the NIR
\citep[e.g.,][]{LadaAdams92,Meyer97,Williams11,Beltran16}. Disks
radiate in a wide range of wavelengths, including in the NIR,
mid-infrared (MIR), and submillimeter. Wide-field photometry in these
wavelengths allows characterizing entire populations and is therefore
extensively used to infer disk demographics in young clusters
\citep[e.g.,][]{Lada87,Lada06}. In this context, the fraction of
sources with disks, their distribution, and the relation with the
stellar density of the cluster and with the mass of its most massive
member(s) can be used to understand how disks are affected by
environmental conditions and to determine the timescales over which
they evolve. Specifically, the presence of strong UV radiation fields
from massive stars and close encounters between objects in a cluster
environment during its early stages can potentially erode, truncate,
or even destroy the circumstellar envelopes and/or disks that feed the
YSO during its formation and that may form planetary systems
\citep[see ][and references therein for a comprehensive
review]{Concha-Ramirez2021arXiv}.

The wavelength range that is used to probe disks and disk properties
around YSOs determines the underlying phenomena to which it is sensitive:
the present paper focuses on NIR observations, which probe
the hot dust in the inner disk, whereas longer-wavelength observations probe
the cooler dusty component farther out. According to theory, the
environment should impact the gaseous component more strongly than the
dusty part of the disk because dust grains are thought to grow
rapidly into a mass that is not easily entrained by the outflow of gas
\citep[e.g.,][]{Sellek20,Haworth18a}, and they are expected to affect mostly
the outer parts of disks, which are less bound
\citep[e.g.,][]{Johnstone98,Adams04}. These effects cannot be traced
accurately with NIR observations. Some other
factors should, in principle, impact the excess emission in the
NIR, such as dust grain growth, gap openings in the inner
disk induced by the formation of planetesimals or by photoevaporation,
or deficient replenishment of the inner disk associated with radial
drift of dust and the truncation of the disk at large radii
\citep{Armitage1999,Drake2009,Owen2011,Facchini16,Sellek20}. The
timescales for these processes as a function of environment
are still largely unconstrained by observations.

Observationally, it has been suggested that the environment has a
measurable impact on the disk population of a young cluster
\citep[e.g.,][]{Guarcello2007,Guarcello2009,Guarcello2010,Fang2012,Balog2007,
  Damiani16, Ansdell17,Stolte10,Mesa-Delgado16,Mann14}, and but no
significant differences between disk statistics in different
environments have been reported as well
\citep[e.g.,][]{Roccatagliata11,Barentsen11,Richert15}. These
seemingly inconsistent results can be partially explained by the fact
that different studies use different disk tracers that probe
physically different components of the disks, and that each technique
is subject to different biases, namely from preferential sensitivity
to sources with or without disks or inadequate spatial resolution. In
this paper we focus on the infrared excess in the range $1$ to
$2.5~\mu$m ($J$, $H$ and $K_S$ bands), which again traces (hot) dust
in the inner ring. These wavelengths are usually considered poorer
tracers of disks than MIR observations, but we present a technique
that boosts the sensitivity of these data to circumstellar material
while keeping the advantages of deep, high-resolution observations in
these wavelengths.

\vspace{0.2cm}

The observed photometric properties of a YSO (in the NIR)
are determined by its mass, age, distance, interstellar extinction,
and emission from circumstellar material. With reference to the
intrinsic properties corresponding to a given mass and age, which we
refer to as ``photospheric'' even though the concept of stellar
photosphere is not well defined for YSOs at their earliest phases of
evolution, distance and extinction will dim the fluxes of YSOs,
extinction will further redden their colors, and circumstellar
material will simultaneously cause additional extinction, add flux,
and typically redden the colors of YSOs even further in the NIR (1 to
2.5 $\mu$m) due to the reprocessing of light by the circumstellar
material. We refer to the contribution of the circumstellar material
to the observed flux as ``excess'' or ``excess emission'' for
simplicity. The observed magnitude, $m_\lambda$, of a source in the
band with wavelength $\lambda$ is then

\begin{equation}
  \label{eq:observed_mag}
  m_\lambda=M_\lambda + \mathit{DM} + A_\lambda + e_\lambda,
\end{equation}
 
\noindent where $M_{\lambda}$ is the intrinsic magnitude of the source, and
$\mathit{DM}$, $A_\lambda$ , and $e_\lambda$ are the distance modulus
($DM=5\log_{10}{d}-5$, where $d$ is the distance in parsec), the
extinction, and the excess, respectively, in magnitudes at the
corresponding wavelength. $e_\lambda$ relates to flux units by

\begin{equation}
  \label{eq:excess_def}
  e_\lambda = -2.5 \log(1+r_\lambda),
\end{equation}

\noindent where
$r_\lambda=F_{\lambda,\mathit{excess}}/F_{\lambda,\mathit{int}}$ is
the ratio of the flux from the circumstellar material,
$F_{\lambda,\mathit{excess}}$, and the intrinsic, photospheric
flux of the YSO, $F_{\lambda,\mathit{int}}$. Since $r_\lambda$ is
always positive, $e_\lambda$ is always negative.

Extinction and excess from circumstellar material affect the colors
and magnitudes of YSOs in slightly different ways, but they are
degenerate past a certain point and are therefore impossible to
distinguish completely with photometric NIR data alone. One subsample
of sources for which this is usually not recognized as a problem is a
sample composed of objects that in a $(J-H)$ \vs\ $(H-K_s)$
color-color diagram lie inside the band that is defined by the
projection of the expected photospheric colors along the reddening
vector (hereafter referred to as ``reddening band''). Sources in this
region have colors consistent with reddened photospheres and are
widely assumed to have little or no excess from circumstellar material
for this reason. One of the results of this paper is that this
assumption can lead to significant errors in the estimate of
individual masses for these sources and to an underestimation of the
number of sources with excess.

Conversely, sources that fall redward of the reddening band must
necessarily have excess with respect to their photospheres. This excess
is attributed to the presence of circumstellar material
\citep{LadaAdams92,Meyer97}. For this population of YSOs, both the
extinction and the individual excess are treated as unknowns. While
the effect of extinction as a function of wavelength is relatively
well characterized and relatively universal, the excess is not as
predictable. In the past, several authors tried to model the effect of
excess from circumstellar material and accretion on the
NIR colors and broadband magnitudes of YSOs based either
on models or on observed populations. We briefly summarize these models below.

\citet{LadaAdams92} used disk models to define the characteristic
locations of classical TTauri stars (CTTS) and Herbig AeBe stars in
NIR color-color diagrams.

Based on a sample of classical TTauri stars from Taurus with
spectroscopically derived properties, \citet{Meyer97} defined an
empirical locus for unreddened CTTS, or sources with substantial
circumstellar disks, in a \ccd. This locus has since been known as the
``CTTS locus'', and it has been extensively used in studies of YSOs
for more than two decades. In the photometric system of our data, it
is defined by
$(J-H)_{\mathrm{CTTS}} = 0.56~(H-K_s) _{\mathrm{CTTS}} + 0.48$. This
locus intersects the main-sequence colors around spectral type M0
\citep{Meyer97}, but the same locus intersects the 1 Myr
pre-main-sequence colors at lower masses, around 0.14
M$_\sun$. According to \citet{Meyer97}, the slope of the CTTS locus
should be valid for the range of masses of TTauri stars, but the
intercept should be a function of spectral type or mass. This means
that the \citet{Meyer97} locus of CTTSs described only a slope for the
displacement of colors in this diagram due to the presence of a
circumstellar disk. The CTTS locus should therefore be regarded as a
vector rather than an actual locus in this diagram, and for this
reason, unless we can determine the extinction independently, it is
impossible to estimate the mass of a source accurately because its
color excess is ill determined. Conversely, because the CTTS locus
only defines a slope, it cannot be used to accurately estimate
extinction without the knowledge of the underlying mass of the object,
which is necessary to determine the intercept of the relation.

\citet{Hillenbrand00} used the same sample of spectroscopically
characterized sources in Taurus as \citet{Meyer97} to derive an
empirical relation between the excess in $K_S$ and in $(H-K_S)$:
$|e_{K_S}|=1.785|e_{(H-K_S)}|+0.134$, with a scatter of $\pm 0.25$
mag. The intercept in this relation must also be a function of mass,
suggesting that this scatter reflects the mass range of the sample, at
least partially. Because they lacked $J$-band observations, the authors
were unable to determine the extinction independently of the excess.
They therefore used this relation and a probability distribution for $|e_{K_S}|$
to infer the distribution of intrinsic colors of YSOs in a sample from
the Orion nebula cluster.

\citet{Lopez-Chico07} proposed another set of relations to
characterize the excess. These authors integrated SEDs from existing
disk+accretion models of \citet{DAlessio98,DAlessio99,DAlessio01} in
the \ccd\ and in the $J$ \vs\ $(J-K_S)$ color-magnitude diagram. They
reported that disks and the luminosity from accretion shift the colors
and magnitude of YSOs consistently along vectors with slopes $0.288$
and $-1.090$ in a \ccd\ and in a $K$ \vs\ $(J-K)$ color-magnitude
diagram, respectively, for stellar masses from 0.8 to 2.3 M$_\odot$
and for a specific set of disk and accretion parameters. They used
models for a fixed age of 10 Myr and for a fixed disk inclination of
$30^{\circ}$. The authors did not assess the impact of these
restrictions, in particular, the consequences of applying the results
from models for 10 Myr objects to younger objects, which are typically
more active in accretion and have denser disks. The slope in the \ccd\
they derived is significantly different from the CTTS locus slope of
\citet{Meyer97a} (which is 0.58 in their photometric system), and the
authors estimated an average difference of 12\% in the estimate of
mass when using the latter with respect to the slope they derived.

In a study of the RCW 108 cluster, \citet{Comeron07} described the
excess using a vector with slope $|e_H|/|e_{K_S}|=0.39$, which they
derived from the models of \citet{LadaAdams92} as the average of
the corresponding slope for CTTS ($0.36$) and that for Herbig Ae/Be
stars ($0.43$), assuming furthermore that the excess in the $J$ band is
negligible. In this notation, since
$|e_{(H-{K_S})}| = |e_{K_S}| - |e_H|$ (because $e_\lambda$ is negative
and $|e_{K_S}| \ge |e_H|$), the slope of the relation empirically
derived by \citet{Hillenbrand00} is $0.44$ for this color-magnitude
combination. This value is comparable to the value quoted by
\citet{Comeron07}, but for the models of Herbig AeBe stars rather than
CTTS.
  
These models provide clever ways to separate and quantify
independently, at least to a certain point, the two unknowns extinction and excess emission and determine the mass of individual
sources. The knowledge of these disk vectors and a set of
extinction coefficients breaks the degeneracy between extinction and
excess, and allows the determination of unreddened intrinsic colors
that can be compared to pre-main-sequence models to estimate the mass
for each source. However, the existing models for these vectors do not
necessarily agree among themselves, and they all rely on assumptions
whose applicability has not been validated for relevant ranges of
age or mass. This limits their use or the robustness of their
interpretations in the study of young populations.
  
\vspace{0.2cm}

We use NIR data for the very young star cluster RCW 38 to
describe a procedure that maximizes the use of broadband NIR
observations between 1 and 2.5 $\mu$m (bands $J$, $H$, and $K_S$),
considering several combinations of colors and magnitudes and minimal
assumptions regarding their underlying properties. \object{RCW 38} (RA
$8^h59^m05.6^s$, DEC $-47^\circ30^m40.8^s$, J2000.0) is an HII region
in the constellation of Vela that hosts a very deeply embedded young
stellar cluster. The distance toward this region has been determined
using 21 cm maps \citep[1-2 kpc,][]{Radhakrishnan72}, radial
velocities from molecular lines and 21 cm emission \citep[1.5
kpc,][]{Gillespie79}, ZAMS fitting of individual sources \citep[1.7
kpc,][]{Muzzio79}, and spectral parallax in B and V \citep[1.7
kpc,][]{Avedisova89}. We adopt a distance of 1.5 kpc to the
cluster. The age of the RCW 38 cluster has been independently
estimated by several authors, all suggesting a value between 0.5 and 1
Myr \citep{Wolk06,Winston11,Getman14}. We adopt an age of 1 Myr for
the cluster.

Massive clusters are rare within reasonable distances, but studying them
is paramount for understanding the process of star formation in all
types of environments. Massive clusters generally have higher stellar
densities and contain massive YSOs alongside low-mass YSOs, allowing
access to the product of star formation in extreme environments. RCW
38 is one of the few massive young clusters within 2 kpc of the Sun,
and its young age makes it particularly relevant for studies of star
formation. One of the goals of this paper is to characterize the
fraction of sources with circumstellar disks in one of these
environments.

The highest-mass object in the cluster is \object{[FP74] IRS 2} (RA
$08^h59^m05.5^s$, DEC $-47^\circ30'39.4"$, J2000). It was initially
classified as an emission peak at 2.2 $\mu$m \citep{Frogel74} and was
later identified as a binary star consistent with spectral types O4 to
O5.5 \citep{DeRose09}.

\vspace{0.2cm}

This paper is organized in the following way: in section
\ref{sec:modelsanddata} we present the models we use for YSO masses
and luminosities, the extinction law we adopt, and our NIR dataset. In
section \ref{sec:red-flag} we identify a problem when the data
are used only partially to derive physical parameters for YSOs. In section
\ref{sec:allowing-excesses} we propose a method that guarantees that all
available data are used consistently. This enhances the use of
NIR data to independently estimate distances, membership,
cluster excess fractions, and constrain individual masses. We present
the results of this method as applied to RCW 38 in section
\ref{sec:results} and consider its adequacy to other star-forming
regions. Finally, we present the conclusions of this study in section
\ref{sec:conclusions}.

\section{Models and observations}
\label{sec:modelsanddata}

In this section we describe the models and dataset used throughout the
paper.

\subsection{Pre-main-sequence models and extinction law}
\label{sec:models}

We used the evolutionary models of \citet{Baraffe15} for masses below
1 M$_\sun$ and the PARSEC v1.2S models
\citep{Bressan12}\footnote{\url{http://stev.oapd.inaf.it/cgi-bin/cmd}}
for masses above this value. We converted the colors and magnitudes of the models
from the 2MASS photometric system into that of our data (SOFI/ISAAC)
using the following set of equations, derived from \citet{Carpenter01}
and from the SOFI instrument
description\footnote{\url{https://www.eso.org/sci/facilities/lasilla/instruments/sofi/inst/setup/Zero_Point.html}}:

\begin{align}
  \label{eq:color_equations}
  J &= J_2 + 0.002~(J-K_s)_2 + 0.008  \nonumber \\
  H &= H_2 + 0.025~(J-H)_2 + 0.006~(H-K_s)_2 + 0.005 \nonumber \\
  K_s &=  K_{s,2} - 0.022~(J-K_s)_2 + 0.010. 
\end{align}

We adopted the extinction law of \citet{Rieke85}, for which
$A_K/A_V=0.112$, $A_H/A_V=0.175$, and $A_J/A_V=0.282$. The extinction
law is accepted as relatively universal in the NIR
\citep[e.g.,][]{Indebetouw05}, although a few studies have argued for
a fair range of extinction coefficients. The distribution of data
points in the \ccd\ rules out extinction laws that produce extinction
vectors with slopes that different strongly from $\sim$~1.7. We
confirmed this with the current dataset and with a wider sample from
2MASS to include more background sources. \citet{Cardelli89} (slope
1.2) and the $\alpha$ value of \citet{Gonzalez-Fernandez14} (slope
2.0) are inconsistent with the data, and the
\citet{Gonzalez-Fernandez14} $\beta$ value (slope 1.9) is still only
marginally consistent. The extinction laws from \citet{Indebetouw05}
(slope 1.73) and from \citet{Nishiyama09} (slope 1.74) are consistent
with the data, and we discuss the impact of using these coefficients
in the determination of the excess fraction of the cluster in
Sect. \ref{sec:excess-fraction}.

\subsection{Observations}
\label{sec:data}

Our dataset is composed of observations in the NIR with SOFI
\citep[ESO/NTT,][]{Moorwood98_SOFI} and in the $J$ (1.25 $\mu$m), $H$
(1.65 $\mu$m), and $K_s$ (2.16 $\mu$m) broadband filters, with ISAAC
\citep[ESO/VLT,][]{Moorwood98_ISAAC} and the $J$ (1.25 $\mu$m), $H$
(1.65 $\mu$m), and $K_s$ (2.16 $\mu$m) filters, and with NACO
\citep[ESO/VLT,][]{Lenzen03,Rousset03} in the $J$ (1.265 $\mu$m), $H$
(1.66 $\mu$m), and $K_s$ (2.18 $\mu$m) broadband filters.

The SOFI data were collected on the night of March 17, 2005 (ESO
program 074.C-0728(A)), using two different instrumental setups for
long and short exposures for increased dynamic range. The long
exposures were taken in large-field mode (field of view of
$4\farcm9 \times 4\farcm9$, pixel scale $0\farcs288$/pixel) to image
the largest possible area of the cluster, and the short exposures were
taken in small-field mode (field of view of
$2\farcm4 \times 2\farcm4$, $0\farcs144$/pixel) to image the inner and
most highly concentrated part of the cluster. This mode is less
sensitive, therefore the saturation of the brightest stars that are
mostly located in the central part of the cluster is minimized.

The ISAAC data were collected on the night of January 18, 2003 (ESO
program 70.C-0729(A)). The cluster was imaged in {\tt on-off} mode
with {\tt jitter} and unit exposure time of 3.5 s in $J$, $H$, and
$K_s$. The field of view for these exposures is $2.5' \times 2.5'$
(Hawaii arm, $\sim0\farcs1484$/pixel).

The NACO {\it JHK$_s$} data were taken on the night of February 23,
2003 (ESO program 70.C-0400(A)), using the S54 camera (field of view
$56''$, 54.3 mas/pixel). The cluster was imaged in {\tt on-off} mode
with {\tt jitter} and unit exposure times of 4.0, 0.75, and 0.5 s to compensate for the heavy extinction that affects the
shorter wavelengths more strongly.

\subsubsection{Data reduction} \label{sec:datared-nir}

The NIR data were reduced with IRAF \citep{Tody86,Tody93}, and the subimages were registered using the {\tt ECLIPSE jitter}
\citep{Devillard97} pipeline and IRAF {\tt xdshifts}.

The reduction of the SOFI and NACO data followed the standard
procedure: crosstalk correction, flatfield, sky subtraction,
registering, and averaging of the subframes. The reduction of the
ISAAC data was all standard, except for the sky subtraction, which had
to be adjusted to accommodate the insufficient number of {\tt off}
frames: the stars in the sky frames were subtracted before
the images were combined to produce the master sky frame.

\begin{figure} 
  \includegraphics{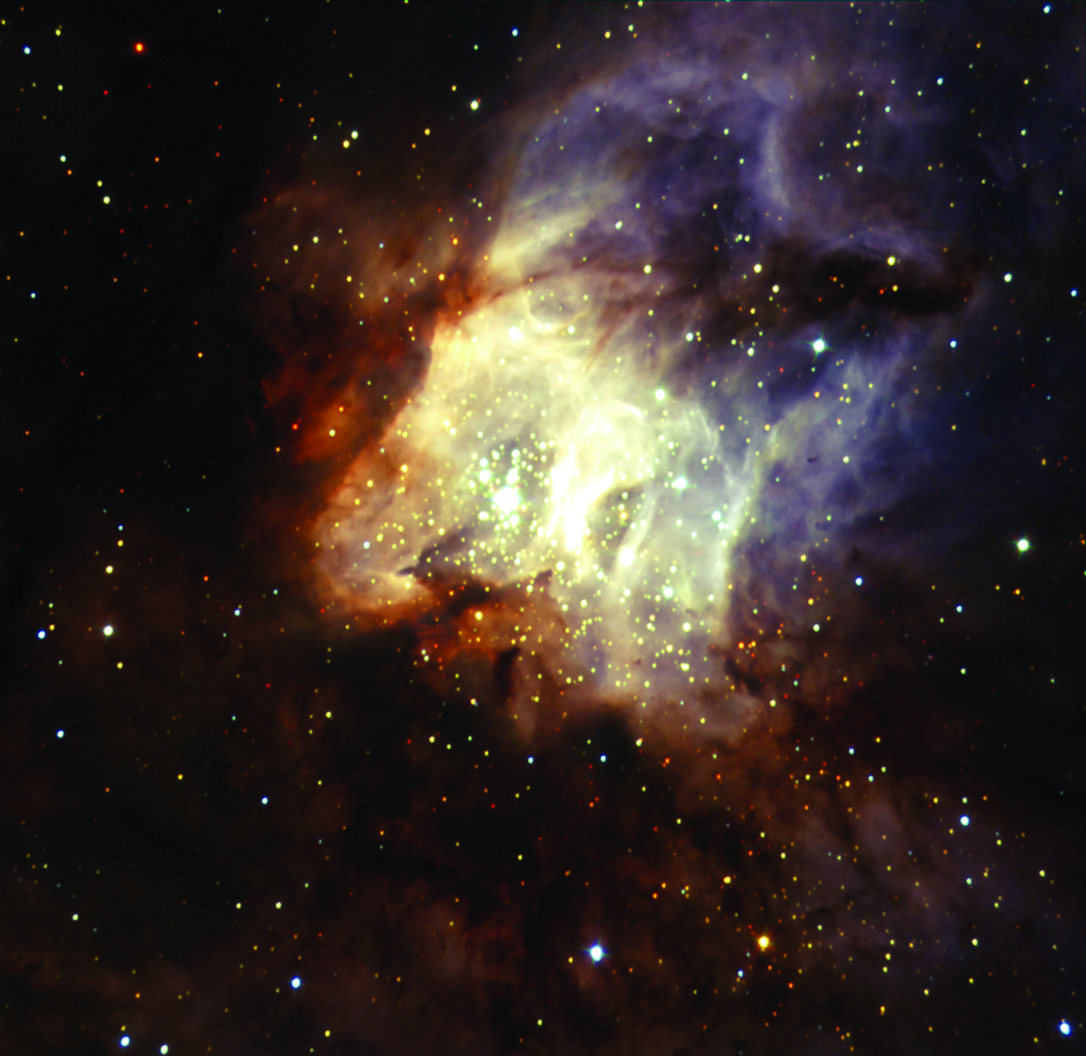}
  \caption{{\it JHK$_s$} color-composite of RCW38 imaged with the
    SOFI instrument, which has a wider field of view in our
    dataset. Equatorial north is up and east is to the left; red is
    $K_s$, green is $H,$ and blue is $J$. The field of view covers
    $4.1' \times 3.9'$ in longitude and latitude    ($1.8 \times 1.7$ pc at a distance of 1.5 kpc).}
\label{fig:rcw38-3colour}
\end{figure} 

The area of the final images for photometry was
$2\farcm12 \times 1\farcm79$ for ISAAC, $4\farcm07 \times 4\farcm19$
for SOFI long exposures, $2\farcm43 \times 2\farcm22$ for SOFI short
exposures, and $47\arcsec \times 44\arcsec$ for NACO {\it JHK$_s$}. 

Figure \ref{fig:rcw38-3colour} shows the color image of the cluster
composed of the three final SOFI long-exposure frames in $K_{s} $
(red), $H$ (green), and $J$ (blue).

\subsubsection{Source extraction and
  photometry} \label{sec:photometry-nir}

Source extraction and photometry were made using DAOPHOT under IRAF
v.2.16.1 for the SOFI data, and with the standalone version of {\it
  DAOPHOT II} (P. Stetson, private communication) for the ISAAC and
NACO data. Astromatic/SCAMP and SWARP \citep{Bertin06,Bertin02} were
used to correct for the field distortions on the ISAAC data and to
calibrate the astrometry of all images.

Table \ref{tab:daofind} lists the main detection parameters. All
detections were made to a $4$ to $5\sigma$ level.

\vspace{0.5cm}

\begin{table}[!h]
  \caption{Parameters for the source extraction of the {\it JH$K_s$}
    data.}
\label{tab:daofind}
\centering
  \begin{tabular}{c c c c}
    \hline\hline
    Dataset & FWHM (pixels) & FWHM (arcsec) \\
    \hline
    SOFI long exp. & $2.1 - 2.6$ & $0.60 - 0.75$ \\
    SOFI short exp. & $5.0 - 6.7$ & $0.72 - 0.96$ \\
    ISAAC {\it JHK$_s$} & $2.7 - 3.2$ & $0.40 - 0.47$ \\
    NACO {\it JHK$_s$} & $2.0 - 4.0$ & $0.1 - 0.2$ \\
    \hline 
\end{tabular} 
\end{table}

We performed point spread function (PSF) photometry on the stars of
the cluster because there was crowding and/or because the PSFs showed
significant distortions, especially in the NACO data due to the
anisoplanatism from the adaptive-optics correction. The last version
of DAOPHOT II allows fitting a third-order polynomial to the
residuals of the analytic PSF fitting, making it ideally suited for
data with a strongly varying PSF across the field of view. Only sources
with photometric errors smaller than 0.15 magnitudes were kept.

The {\it JHK$_s$} magnitudes of isolated stars in common for the SOFI long
exposures and 2MASS were compared to obtain the instrumental zeropoint
using the median, after accounting for the difference in photometric
systems. These zeropoints were then used to cross-calibrate the
magnitude scale of the short exposure images and of the ISAAC data by
comparing the magnitudes of the stars in common. ISAAC calibrated
photometry was then used to cross-calibrate the NACO data. The
instrumental zeropoints calculated in this way are summarized in Table
\ref{zp}.
 
\vspace{0.5cm}

\begin{table}[h]
 \centering
  \begin{tabular}{c c c c}
    \hline\hline 
    & Band & ZP$_{inst}$ (mag) & rms (mag)\\ 
    \hline 
   & $J$ & 23.056 & 0.041 \\
    SOFI Long exp. & $H$ & 23.009 & 0.079 \\
    & $K_{s}$ & 22.263 & 0.121 \\
    \hline
    & $J$ & 22.933 & 0.055 \\ 
    SOFI Short exp. & $H$ & 22.812 & 0.043 \\
    & $K_{s}$ & 22.003 & 0.061 \\
    \hline
    & $J$ & 15.007 & 0.235 \\
    ISAAC & $H$ & 14.079 & 0.220 \\
    & $K_{s}$ & 13.822 & 0.198 \\
    \hline
    & $J$ & 15.563 & 0.260 \\
    NACO & $H$ & 17.23 & 0.201 \\
    & $K_{s}$ & 20.286 & 0.192 \\
    \hline 
  \end{tabular}
  \caption{Instrumental zeropoints for the science frames.}
 \label{zp}
\end{table}

\subsubsection{Final sample}
\label{sec:final-sample}

\begin{figure*}
\centering
  \includegraphics[width=19cm]{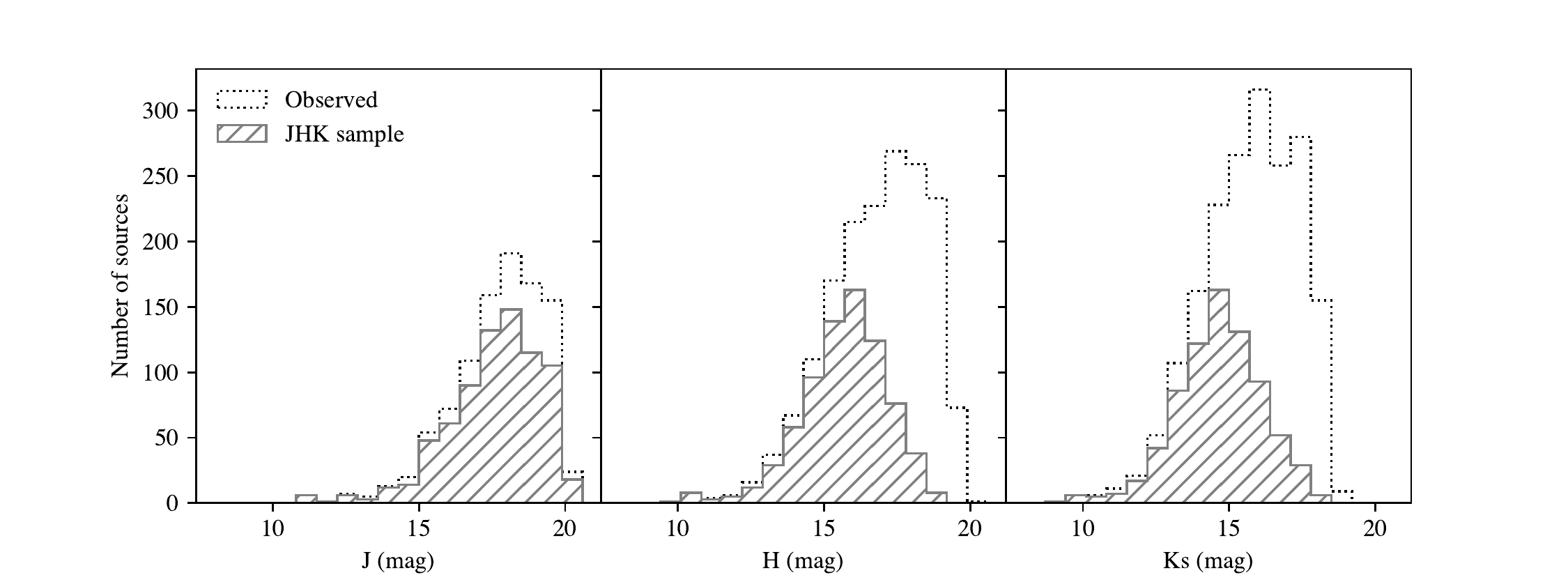}
  \caption{Distribution of the apparent brightness of all observed sources
    ({\it dotted histograms}) and of the sources detected concurrently
    in $J$, $H$, and $K_S$ (final sample; {\it hatched histograms}).}
  \label{fig:jhk_data}
\end{figure*}

\begin{figure}
  \centering
  \resizebox{\hsize}{!}{\includegraphics{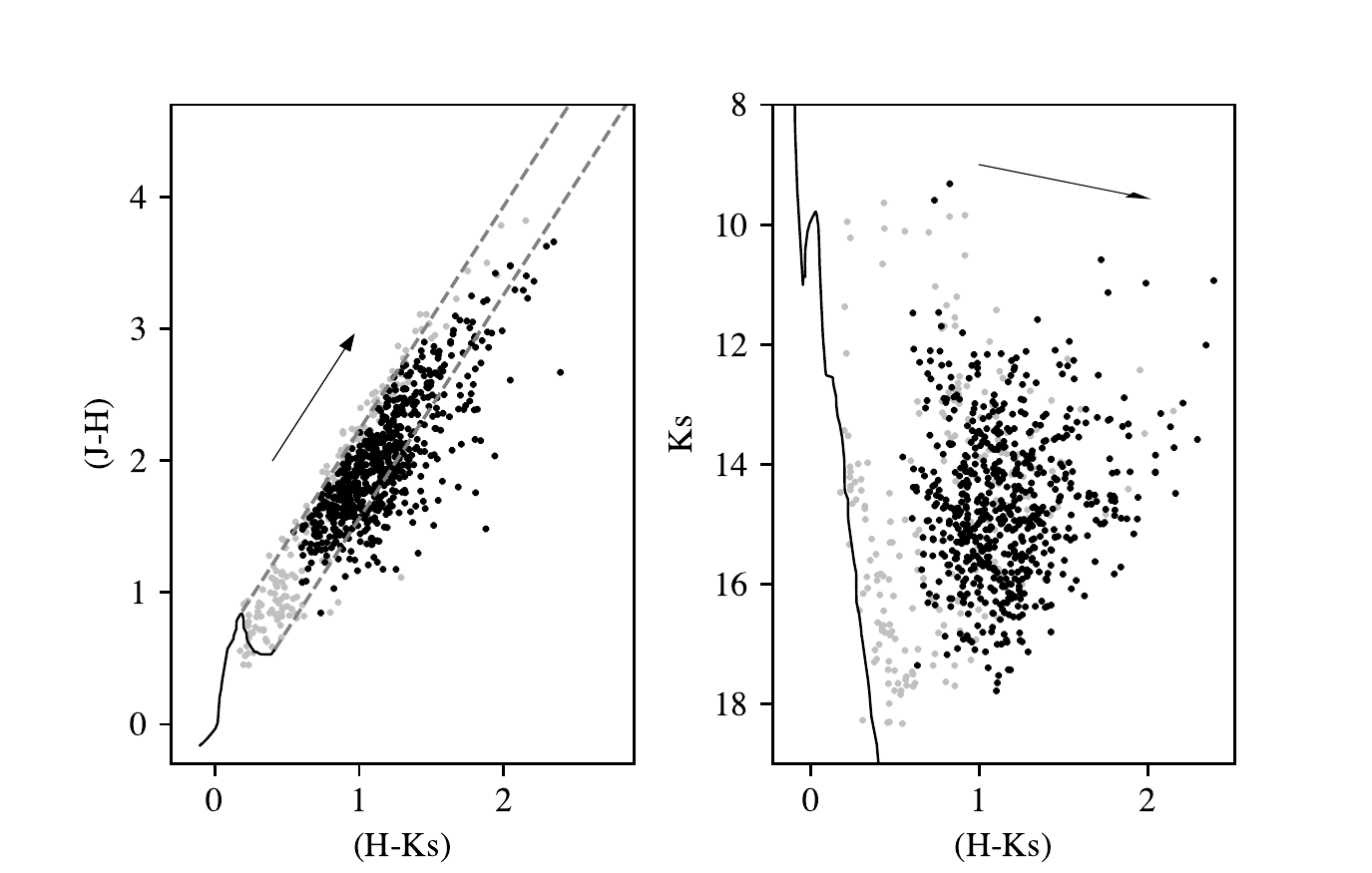}}
  \caption{\ccd~and $K_S$ \vs\ $(H-K_S)$ color-magnitude diagram for
    the observed sources. The {\it black symbols} correspond to the
    sources that are probably cluster members (see
    Sect. \ref{sec:contaminants}), and the {\it gray symbols}
    correspond to the remaining sources in the field of view
    (contaminants). The {\it solid lines} in both diagrams represent
    the adopted models for an age of 1 Myr and a distance of 1.5
    kpc. The {\it arrows} show the displacement caused by an
    extinction of 1 magnitude in $K_S$ according to the assumed
    extinction law (see Sect. \ref{sec:models}). The {\it dashed
      lines} show the limits of the reddening band in the color-color
    diagram, defined by the reddening vector and the extreme ends of
    the model isochrones in color.}
  \label{fig:cc-cmd}
\end{figure}

The NACO, ISAAC, and SOFI catalogs were merged into a master catalog
in which only the sources with the smallest photometric error in the
overlap regions were kept. In general, this amounts to keeping all sources from
NACO (better resolution), except for sources in the distorted corners,
all sources from ISAAC outside the NACO field of view, and all sources
from SOFI outside the ISAAC field of view, plus a few inside
corresponding to sources that fell on cosmetic defects of the ISAAC
detector or to sources saturated in ISAAC (and NACO). The final master
catalog contains 2102 sources, 985, 1697, and 1879 of which are detected in
$J$, $H$, and $K_s$, respectively. 1560 and 875 sources are detected
concurrently in $H$ and $K_s$, in $J$, $H$ and $K_s$, respectively. Of
the latter, the colors of 21 sources lie too far to the left of the
blue limit of the reddening band in the \ccd, which is suggestive of poor
photometry (e.g., blended sources or bright nebula emission),
and were eliminated from the dataset. We did not
use the sources that were detected only in $H$ or/and in $K_S$. The
final sample therefore consists of 854 sources with photometry in $J$,
$H,$ and $K_S$. Figure \ref{fig:jhk_data} shows the distribution of
apparent magnitudes in the initial and in the final samples. Of the
854 sources, 729 lie inside the reddening band in a \ccd, and 125 lie
to the right of the reddening band.

We show below that the method we propose naturally distinguishes
between cluster members and unrelated sources. We therefore did not perform
any YSO selection prior to applying the method.

\section{Estimating parameters from NIR data: The problem}
\label{sec:red-flag}

In Section \ref{sec:allowing-excesses} we present a new
approach to using NIR photometric data, but this approach was born
from the realization that using the data only partially, as is common
practice, may lead to incorrect interpretations. In this section we
describe this problem.

We attempted to estimate the intrinsic properties of our sources, that is,
their masses and intrinsic luminosities, by comparing their observed
fluxes and colors to existing models using standard procedures.

We considered as before that the observed properties of each source
may be affected by distance (via the distance modulus equation), by
interstellar extinction (both foreground extinction and extinction from the molecular
cloud itself), and by emission in excess of their photospheric
emission from circumstellar material (see
Sect. \ref{sec:introduction}). However, for now, we focus only on the
subset of sources inside the reddening band (see
Sect. \ref{sec:final-sample}). These are typically assumed to have
little or no excess emission from circumstellar material because their
colors are consistent with reddened photospheres. We follow this
assumption at first and then demonstrate that it leads to
inconsistent results.

\subsection{ \ccd}
\label{sec:redband-ccd}

\begin{figure*}
\centering
  \includegraphics[width=19cm]{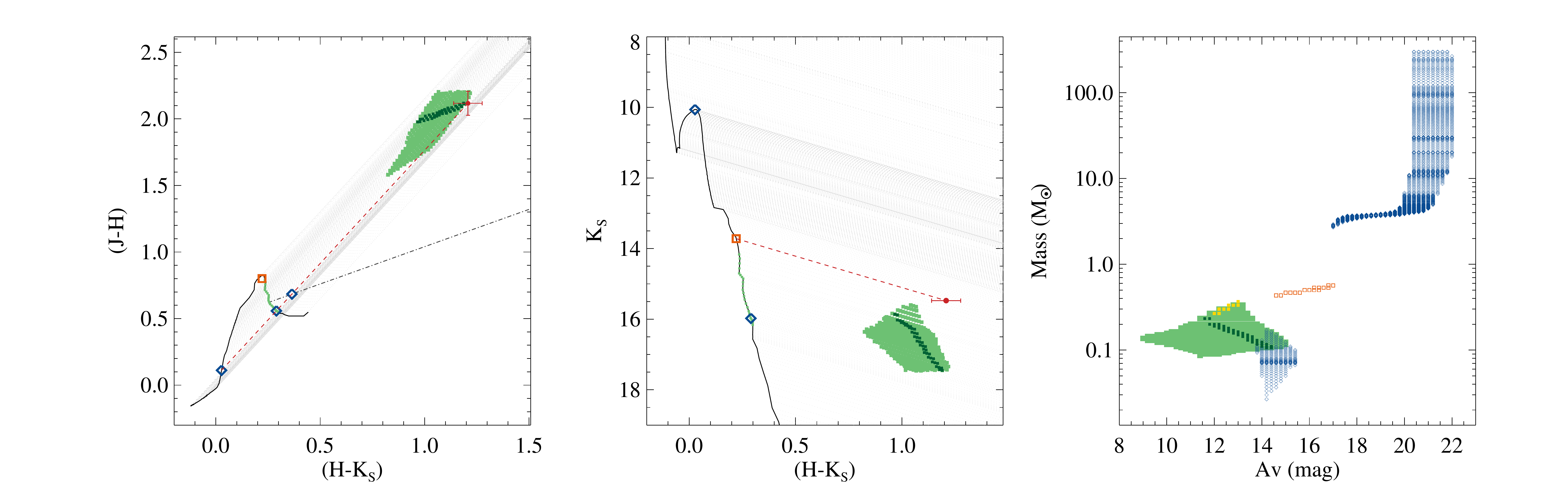}
  \caption{Estimating physical parameters in color-color and
    color-magnitude space ({\it left} and {\it middle} panels,
    respectively) for one example source inside the reddening
    band. The {\it solid black lines} show the pre-main-sequence
    evolutionary models in the two panels, and the {\it dot-dashed
      black line} shows the canonical CTTS locus of \citet{Meyer97} in
    the color-color diagram. The {\it filled red circles} and error
    bars mark the observed colors and $K_S$ magnitude of the source
    and its photometric errors in the two panels. The {\it dashed red
      lines} show the projection of the observed source toward the
    models along the appropriate reddening vectors for each panel.
    These lines intersect the models at the points marked with the
    {\it blue diamonds} in the CC diagram and at the points marked
    with the {\it orange square} in the CMD. The corresponding model
    points are marked with the same symbols in the two panels, showing
    that the two estimates do not overlap. The figure also shows the
    grid of reddened model points ({\it light gray dots}) described in
    Sect. \ref{sec:allowing-excesses}. The {\it green areas} limit the
    range of reddened colors and magnitude that provides consistency
    across all diagrams by allowing for the presence of excess
    emission (see sect. \ref{sec:allowing-excesses} for details). The
    {\it right panel} shows the corresponding ranges of mass and
    extinction for this example source. For all panels, the models are
    for an age of 1 Myr and a distance of 1.7 kpc.}
  \label{fig:method_redband}
\end{figure*}

We can estimate intrinsic properties from the \ccd\ (CC diagram) only
for sources without excess emission because we do not have enough
information to quantify the excess emission independently. As
mentioned above, we assumed that the YSOs that lie inside the
reddening band in this diagram have no excess emission.

Under this assumption, estimating the mass of each source using the
\ccd, for instance, consists of dereddening the observed colors to the
models along the reddening vector to find the intrinsic colors. As
shown in Fig. \ref{fig:method_redband} (left panel) for one example
source, the distance between the source ({\it filled red circle}) and
the models ({\it solid black line}) along the reddening vector was
measured and converted into a value of extinction. In addition to the
extinction, this returns the intrinsic color(s) ({\it blue diamonds})
for each source. Because of the shape of the models in this diagram,
the reddening line of any one source typically intersects the models
more than once, producing more than one possible pair of extinction
and mass for that source if we use this diagnostic alone (see
figure). We refer to the extinctions and masses determined in this way
as $A_{V,\mathit{CC}}$ and $M_{\mathit{CC}}$, respectively
($\mathit{CC}$ for color-color diagram).

\subsection{Color-magnitude diagram}
\label{sec:redband-cmd}

We can similarly use color-magnitude diagrams (CMDs) for independent
estimates of extinction and mass, again only for sources without
excess emission. If we maintain the assumption that the sources inside
the reddening band in a \ccd\ do not have excess emission, their
intrinsic colors and magnitudes can be found by dereddening the
sources in color-magnitude space to the models, previously shifted
vertically by the distance modulus of the cluster. As shown in
Fig. \ref{fig:method_redband} (middle panel), the distance between the
source ({\it red symbol}) and the models along the reddening vector
provides a measure for extinction, and the point at which this line
intersects the models returns the intrinsic magnitude and color of the
source ({\it orange square}), hence its mass.

This approach is less likely to produce degenerate masses (and
extinctions) than the previous because in color-magnitude space, the
models are degenerate for smaller intervals of mass. Still, in $K_s$
\vs\ \hk\ color-magnitude space, for example, models for an age of 1
Myr are degenerate in this sense between $\approx$2.7 and $\approx$10
M$_\sun$. This way of estimating the mass depends critically on the
knowledge of the distance because the apparent magnitude relates to
the absolute magnitude via the distance modulus equation.

We refer to the estimate of extinction and mass in a $K_s$ \vs\
\hk\ color-magnitude diagram as $A_{V,\mathit{CMD}}$ and
$M_{\mathit{CMD}}$, respectively.

\subsection{Inconsistency between the CMD and the CC diagram}
\label{sec:inconsistency_cc_cmd}

For the current subset of sources (inside the reddening band), we
compared the properties estimated from the $K_S$ \vs\ \hk\ CMD with
those estimated from the color-color diagram under the assumption that
they had no excess. The two estimates should agree if our
assumptions were true, but we find that they are consistent only for
269 (37\%) out of the 729 sources inside the reddening band, even
allowing for photometric errors.

This is illustrated in Fig. \ref{fig:method_redband}: the same source
dereddens to the {\it blue diamonds} in the color-color diagram ({\it
  left panel}) and to the {\it orange square} in the color-magnitude
diagram ({\it middle panel}), corresponding to masses (and
luminosities) that are irreconcilable within the errors. If we compare
$M_{\mathit{CC}}$ with the mass estimated from $J$ \vs\ $(J-H),$ which
is the combination of color and magnitude that should be least
sensitive to excess, we find that the two estimates are still
inconsistent for 40\% of the sources.

This clearly shows that neither estimate is reliable. It also suggests that our
assumptions may be incorrect. We address the possible causes for
these inconsistencies in Appendix
\ref{sec:understanding-discrepancy} .

\section{New method: Allowing for excess}
\label{sec:allowing-excesses}

In the previous section (Sect. \ref{sec:red-flag}), we assumed that
the sources inside the reddening band in a \ccd\ did not have excess
emission from circumstellar material. This is a common assumption
because their colors are consistent with reddened photospheres. We
found a strong inconsistency between their properties as estimated in
the color-color or in the color-magnitude spaces. It is not
unreasonable for these sources to have excess, however: the excess may
be small enough for the sources to not cross the red border of the
reddening band, and/or some sources may have (small) excess in the
flux that is gray and therefore do not alter their colors
significantly. If these sources do have excess, then assuming
otherwise will naturally produce a discrepancy between the CC and CMD
mass estimates because the observed colors and magnitudes in this case
are not photospheric. Moreover, the discrepancy will present as a
scatter rather than a systematic offset because the excess emission
intrinsically depends on many factors.

Because we ruled out other plausible contributions for this
inconsistency (cf. Appendix \ref{sec:understanding-discrepancy}), we
proceed to investigate how allowing for the presence of excess
emission impacts the characterization of these sources.

\subsection{Defining the parameter space}
\label{sec:define-param-space}

In the following we assume one distance (1.7 kpc) and one single age
(1 Myr) for the cluster. We then attempt to constrain the underlying
stellar parameters by finding the range of masses, extinctions, and
amounts of excess emission for which the CC diagram and the CMDs
produce consistent agreement with the models for each source. We
consider that the following assumptions are reasonable:

\begin{enumerate}

\item The excess, if it exists, is always red, that is, the
  observed color of a source cannot be bluer than its photosphere, and
   the excess $e_{\mathit{HK}}$ in \hk\ is always greater than or
  equal to the excess $e_{\mathit{JH}}$ in \jh
  . \label{item:red}

\item The colors of the source after removing the excess are bound by
  the reddening band. This means that the only two phenomena affecting
  the colors of each source relative to their intrinsic colors are
  assumed to be the excess emission and
  extinction. \label{item:redband}

\item The luminosity produced by the circumstellar material cannot be
  greater than five times the flux of the source in the $K_s$ band
  ($r_K \le 5$). This is a conservative estimate for sources that are
  past the protostellar stage for observations in the
  NIR. According to eq. \eqref{eq:excess_def}, this limits the
  excess in $K_s$ to $|e_{\mathit{K_s}}|=1.95$ magnitudes (but see
  sect. \ref{sec:distr-excess}). For reference, the maximum
  excess observed in $K$ in Taurus is 1.6 mag
  \citep{Meyer97a}. \label{item:maxek}

\item The excess in $J$ must be smaller than or equal to that in $H$,
  and the excess in $H$ must be smaller than or equal to that in
  $K_s$ ($|e_J| \le |e_H| \le |e_{\mathit{K_s}}|$). This assumes the
  spectral energy distribution of the excess emission increases from 1
  to 3 $\mu$m. Considering the previous assumption, this also limits
  the excess in $J$ and $H$ to 1.95 magnitudes. \label{item:ejh}
\end{enumerate}

We neglect the contribution of a high accretion luminosity in defining
our set of constraints, in particular by assuming that the excess
emission is red. This could be relevant for the sources with very
high accretion rates, for sources whose accretion flows are exposed,
and for sources whose disks are viewed face-on. We expect most of the
sources in this subsample to have relatively small disks and low
accretion rates already, otherwise they would more likely be found to
the right of the reddening band in the CC diagram.

For this particular cluster, Fig. \ref{fig:cc-cmd} shows that there
are barely any sources at extinctions lower than $A_V \sim 7$ mag,
suggesting that the cluster members are located behind a wall of
extinction likely created by the part of the cloud that lies in
front. We therefore conservatively adopted the additional constraint
that the minimum extinction of any cluster member cannot be lower
than $A_V = 5$ mag.

We note that point \ref{item:maxek} above is probably too permissive because the colors of a source with a disk corresponding to the maximum allowed excess
in $K_S$ would most likely appear to the right of the
reddening band in a \ccd.

We also note that these requirements do not force the existence of
excess, they only allow for it if it is necessary to find consistency
across color and magnitude space. These constraints and this method
therefore answer the  question which excess would be required so that the observed colors and fluxes are consistent with
the models we adopt for the photospheres of YSOs.

\subsection{Method at work}
\label{sec:method}

The above conditions, plus the requirement that all observed fluxes
and colors must be consistent with the models, constrain the range of
intrinsic colors and magnitudes that are plausible for each source,
and this then translates into a range of possible combinations of
extinction, excess emission, and mass for each source.

In other words, rather than working from the observations to the
models, we perform the inverse exercise to search for the combinations
within the models that reproduce the observed photometry of each
source consistently. To do this, we created a grid of points consisting
of the luminosities of the model photospheres reddened by steps of 0.2
magnitudes, then applied the constraints to the grid from the observed
colors and magnitudes of each source and their photometric errors, and
traced back the allowed ranges of mass, extinction, and excess emission.

Figure \ref{fig:method_redband} illustrates the procedure for one
source inside the reddening band. The {\it blue diamonds} and the {\it
  orange square} mark the intrinsic colors and $K_S$ magnitude
expected if the source is dereddened to the models in the CC diagram
and in the HK CMD, respectively, assuming no excess emission. For this
source, the corresponding mass estimated from the HK CMD is $0.48$
M$_\sun$ and that estimated from the CC diagram is either $0.10$ or
$3.72$ M$_\sun$. The {\it right panel}  uses the
same symbols and color codes to illustrate that the corresponding estimates of mass
and extinction are irreconcilable when we assume no excess, even
allowing for photometric errors.

If we allow for excess emission using the set of constraints defined
above (Sect. \ref{sec:define-param-space}), then the {\it green areas}
show the range of de-excessed colors and $K_S$ magnitudes that
guarantee that the CC diagram and the CMDs provide consistent
estimates for this source, and the {\it green portions} of the models
show the corresponding ranges of intrinsic colors and magnitudes. This
source in particular requires an excess in $K_S$ of at least 0.08
magnitudes beyond its photometry uncertainty to produce consistent
estimates across the color and magnitude parameter space. Allowing
for excess emission, the mass of this source is limited to the range
$0.09$ to $0.3$ M$_\sun$, which is 1.6 to 5 times smaller than the
value obtained from the HK CMD assuming no excess.

This example shows that we are exposed to an error that can be large
if we do not allow for excess when the masses are derived from photometric
data, even for the sources inside the reddening band (see
sect. \ref{sec:comparison} for more details). In this case, the
interval of possible masses partially overlaps the mass estimate
from the CC diagram, but this is not the case for all sources.

The example also shows that the ranges of possible properties for each
source are not independent: a fixed extinction, for example, restricts
the interval of possible masses (and excess). Furthermore, the
properties within these intervals are not equally probable, which
means that a Bayesian inference analysis that would incorporate all
information available for each source and more realistic expectations
for the characteristics of circumstellar material could constrain the
properties of each YSO even further. This is beyond the scope of this
paper, however. For example, if we consider that the CTTS locus slope
of \citet{Meyer97} is a good representation of the relation of colors
for YSOs with circumstellar material, then the range of properties
allowed for the example source in Fig. \ref{fig:method_redband} is
limited further to the band defined by the {\it dark green} points;
the mass of this source then becomes restricted to $0.1 - 0.2$
M$_\sun$.

\subsection{Sources to the right of the reddening band}
\label{sec:method-ctts}

The position of sources to the right of the reddening band in the \ccd\ requires that they have
excess emission with respect to their photospheres. In addition
to extinction, we must therefore determine this excess before we can infer their
masses.

\begin{figure*}
\centering
  \includegraphics[width=19cm]{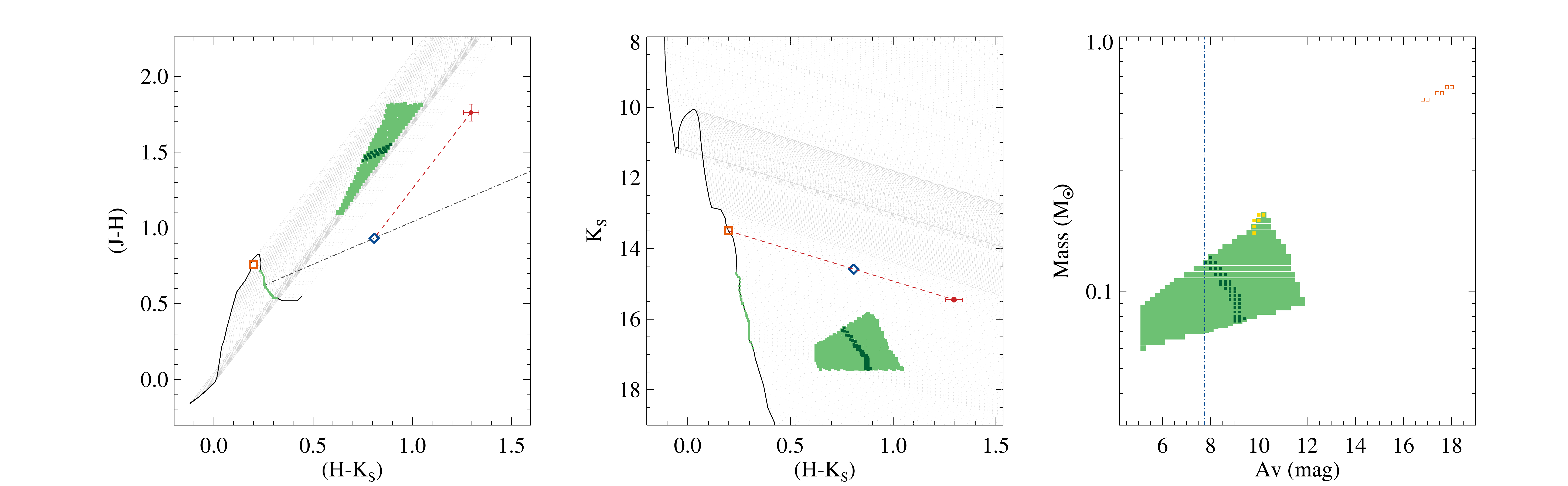}
  \caption{Same as Fig. \ref{fig:method_redband}, but for an example
    source to the right of the reddening band.}
  \label{fig:method_ctts}
\end{figure*}

For these sources, the standard approach relies on taking the CTTS
locus as a single line rather than the direction vector of a continuum
of lines whose intercept varies with the spectral type of each source
(see Sect. \ref{sec:introduction}). This assumption carries an
uncertainty to the extinction that cannot be quantified because the
spectral types are not known \textup{a priori}; although it can be as
high as 1 magnitude in $A_V$ according to \citet{Meyer97}, it is
commonly taken to be smaller than the photometric uncertainty and is
typically ignored. The procedure is then to deredden all sources that
fall within the empirical region of CTTSs in the color-color diagram
to the same one line (the CTTS locus; {\it dot-dashed line} in
Fig.~\ref{fig:method_ctts}) to estimate the extinction of each
source. From the extinction, the only way to estimate excess emission
and then mass is to assume an excess vector, as mentioned in
Sect. \ref{sec:introduction}, in color-magnitude space, but these
solutions have not been extensively adopted by the community.
 
The method we propose to constrain physical parameters, that is, by searching
the parameter space for consistency across color and magnitude space,
can also be applied to these sources. This is indeed currently
the only method that can constrain the mass, extinction, and amount of
excess of any individual source from NIR photometry alone, even though
it cannot constrain these parameters to a single most probable value,
as would be desirable. With other methods, some
more or less arbitrary amount of excess must always be assumed to find a value for the mass
rather than constraining it from the data.

The selection of sources that fall to the right of the reddening
band does not limit the sample to classical TTauri stars, but
includes all sources with excess, that is, Herbig AeBe stars. Our
sample contains 94 such sources in total. These sources are redder
than reddened photospheres and therefore clearly have excess emission,
which activates the condition that their excess must be at least such
that their colors have to move to the reddening band in color-color
space.

Figure \ref{fig:method_ctts} illustrates the procedure for one of
these sources. In this case, the mass estimate obtained by dereddening
directly to the models in an \hk\ \vs\ $K_S$ color-magnitude diagram
(marked in the figure by the {\it orange squares}) is deliberately incorrect because it is clear from the source position in the color-color
diagram that it must have excess, but it is plotted here for
completeness. In the {\it left panel,} the {\it blue diamond} shows
where this source would deredden to in the canonical CTTS locus,
allowing an estimate of its visual extinction, which would be 7.7
magnitudes in this case. As mentioned above, this value of extinction
can be incorrect by more than one magnitude, depending on the actual
underlying mass of the object. In the {\it middle panel,} we show the
position of the source that is dereddened by this amount of extinction (marked
by the {\it  dot-dashed blue line}) in an \hk\ \vs\ $K_S$ CMD. It is
clear that even if we trusted the extinction calculated using the
CTTS locus, we would still lack information to derive the mass of the
source because we do not know its excess. Using our method, we
constrain the mass of the source to 0.06 to 0.2 M$_\sun$ and infer
that its excess in $K_S$ must be greater than 0.4 magnitudes.

\section{Results}
\label{sec:results}

In this section we discuss how this method can distinguish between
YSO candidates and background objects, how it can be used to determine
the distance to equidistant, coeval populations (e.g., young clusters),
and analyze the excess in $K_S$ for the sources in RCW 38.

\subsection{Membership analysis}
\label{sec:contaminants}

The method we propose naturally distinguishes between cluster members
and foreground and background sources, making it a very useful tool in
evaluating membership. The conditions we impose (see
Sect. \ref{sec:define-param-space}) are reasonable for cluster members
because they assume a fixed distance and age and allow a set of
characteristics for excess emission that is plausible for
YSOs. Unrelated sources (or sources with poor photometry) in general
do not comply with these conditions, however, either because their
distance is different from that of the cluster or because their
photometry cannot be reconciled with the models. Therefore, all
sources for which the method cannot find a solution were classified as
unrelated sources and were eliminated from the cluster sample.

In our dataset, we started with 854 sources. The method found a
solution for 705 of these sources. Of these, 123 were below the
reddening band, which means that the method cannot find a solution for
2 sources that lie in the CTTS region of the \ccd; these are sources
that would normally be regarded as cluster members because they
display obvious excess emission, but this method suggests they may
be contaminants, likely extragalactic. The remaining 582 sources for
which the method finds a solution lie inside the reddening band in the
\ccd.

We find that the solution found for a few sources is unrealistic: they
have allowed excess that is very low in color (\textup{}i.e., the
proposed excess is gray in $(H-K_S)$), but high in flux
($|e_{min}|>0.2$ mag). This configuration is unlikely for YSOs because
excess from circumstellar disks is most often red ($|e_K|>|e_H|$),
especially when it is high in flux. Conversely, this presentation is
expected if the sources have a distance different from what is assumed
for the cluster: in this case, the flux in all bands would have the
same excess, which would correspond to the difference in distance
moduli between the cluster distance and the true distance of the
sources. We therefore interpret these sources as background objects
and classify them as unrelated as well. In our dataset, we find 69 of
these additional sources (68 inside the reddening band and one below
the reddening band). This selection does not eliminate all sources
with gray excess in $(H-K_S)$: those that also have low excess in
$K_S$ were kept, as it is physically plausible that these sources
exist, although we admit some of them may still be contaminants. Its
ability to select cluster members means that this method may be useful
in identifying targets for follow-up studies, namely for spectroscopic
surveys.

After the selection above, our science selection contains 636
cluster members (and 218 unrelated sources), which translates into a
membership fraction of 74\% (71\% if we consider only sources inside
the reddening band, as drawn in Fig. \ref{fig:check_distance} for
reference). In the projected area of the field of view, this
amounts to a surface density of about 200 YSOs/pc$^2$ (at a
distance of 1.5 kpc), although we recall that this
dataset is composed of observations using instruments with different
resolutions and sensitivities, and the completeness therefore varies
across the field of view. Incompleteness precludes the detection of
the fainter sources, so that the actual surface density of the cluster is
necessarily higher than the value we determine.

Below we assess the reliability of this method in evaluating
membership.

\subsection{Distance determination}
\label{sec:distance}

As mentioned in the previous section, the conditions we impose with
this method are expected to be reasonable for cluster members provided
that the models are correct\footnote{We did not address the validity
  of the models in this work, but we note that the least excess we
  find is too high to be explained by the models unless they are
  grossly inadequate, which is implausible.} and that the age and
distance are correct. Because we know that this sample contains a
large number of equidistant and probably coeval sources (the cluster
members), the method is expected to result in convergence for the
largest number of sources when the correct distance to the population
is assumed.

\begin{figure}
  \centering
  \resizebox{\hsize}{!}{\includegraphics{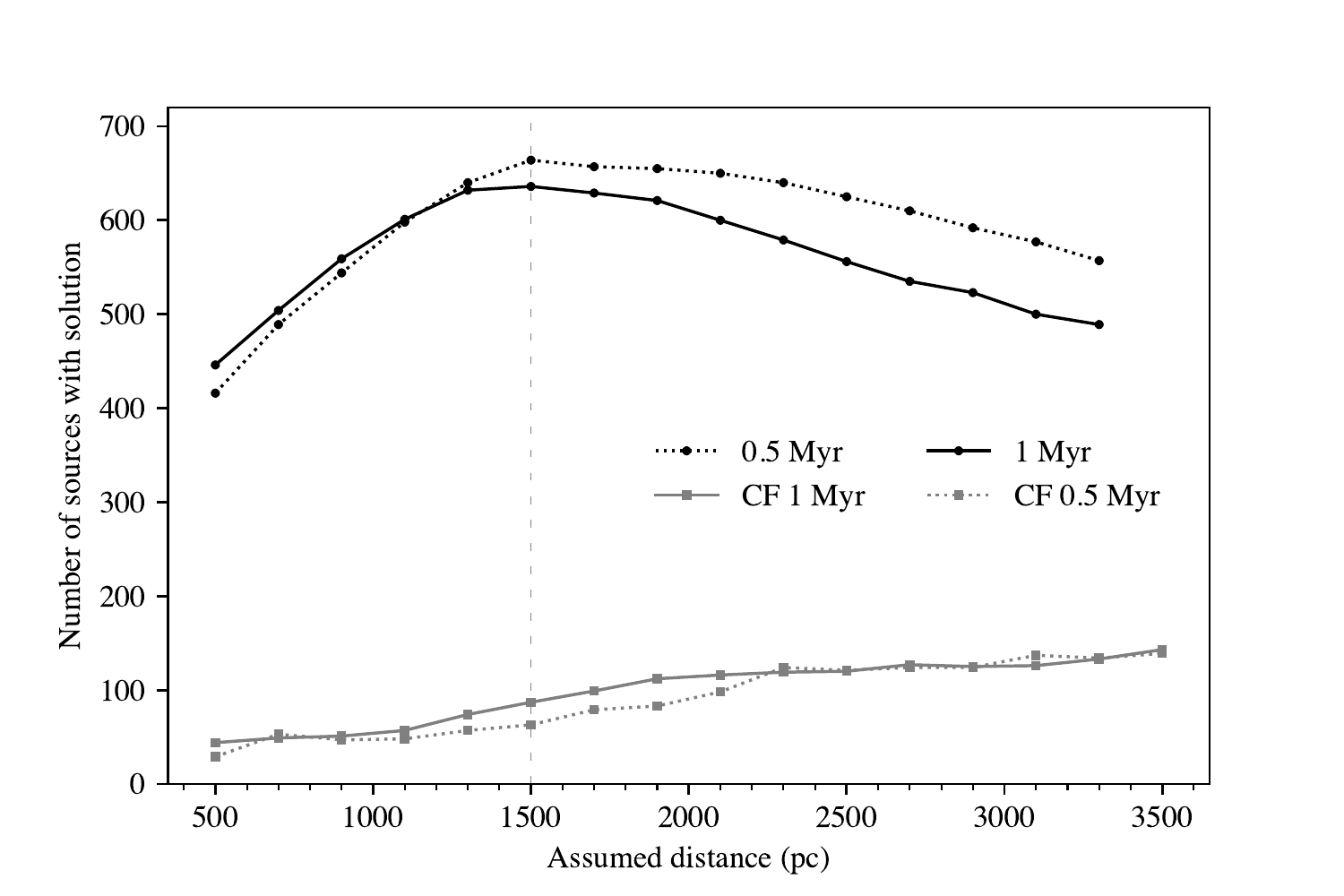}}
  \caption{Number of sources for which the photometry can be
    reconciled with the models in color and magnitude within the
    photometric errors, assuming an age of 1.0 Myr ({\it solid lines})
    or 0.5 Myr ({\it dotted lines}) for the science sample ({\it
      black lines, circles}) and for the control field ({\it
      gray lines, squares}). The vertical {\it dashed line}
    shows the distance we assume for the cluster (1.5 kpc).}
  \label{fig:distance_result}
\end{figure}

We therefore calculated the number of sources with a solution from the
method while varying the assumed distance from 500 to 3500
pc. Fig. \ref{fig:distance_result} shows a distribution that has a
peak, as expected if an equidistant population is present. The peak is
located at $1.5\pm 0.2$ kpc, although, considering its width and the
statistical error, we cannot rule out distances up to 1.9 kpc.  These
values agree well with the distance determined independently in other
works (see Sect. \ref{sec:introduction}), validating this method as a
reliable tool for determining distances to locally dense
populations. Although we did not test this in this work, it is
foreseeable that this method is most accurate for denser
populations without significant age spreads.

For comparison, we applied the same procedure to a nearby control
field taken from the 2MASS Point Source Catalog \citep{Skrutskie06}
from a region at the same Galactic latitude as RCW 38, 1.5$^{\circ}$
West from the cluster in Galactic coordinates. We selected 11,223
sources with good photometry in $J$, $H,$ and $K_S$ and plotted the
number of these for which the method found a solution assuming
distances in the same range. The results are shown as {\it gray
  lines} in Fig. \ref{fig:distance_result}. Although the control
field has 13 times more sources than the science field, the method can
only find a solution for 50 to 150 sources, and the distribution of this
number with an assumed distance does not show any peak.

We can use the control field distribution to estimate how effective
the method is at evaluating membership (see
Sect. \ref{sec:contaminants}). Because we know that the control field is
unlikely to contain sources from the cluster, we can estimate the
number of unrelated sources that the method would erroneously consider
cluster members using the criteria defined in
Sect. \ref{sec:contaminants}. Applying a scaling factor to account for
the difference in area between the control and the science fields, we
estimate that the method would include 0.4 unrelated sources in the science
sample. On the other hand, when we apply a scaling factor that accounts
for the difference in the number of sources up to $K_S=14$ mag (the
limiting magnitude of the 2MASS sample), we estimate that the method would
include 1.3 unrelated sources in the science sample in this magnitude
range. The actual difference in sensitivity between the two samples
precludes a definitive analysis of the contamination that may still be
included in the science sample, but these calculations provide
confidence that the method does not include many contaminants
in the science sample.

Fig. \ref{fig:distance_result} exposes a less obvious feature of this
method: In the science sample, the method still finds convergence for
many sources when assuming the incorrect distance, whereas in the
control field, the method only finds convergence for a residual
fraction of the stars. This happens because most of the YSOs in this
field need excess to reconcile their photometry with the models. If
the assumed distance is not their true distance, then the method may
still find a solution because the difference in distance modulus can
mimic true excess up to some point. Therefore, it is expected that a
sample containing a large number of sources with intrinsic excess
presents a higher baseline in Fig. \ref{fig:distance_result} ({\it
  black lines}), as is observed. The numbers in the wings of this
distribution are very close to the number of sources for which the
method detects excess emission, corroborating that the reason for the
much higher numbers in the science sample is the presence of intrinsic
excess. The photometry of field stars, on the other hand, is
consistent with some model without invoking excess, so that only
sources for which a difference in distance modulus can mimic true
excess when assuming models for an incorrect age will have a solution,
and it is expected that this represents only a residual amount. As
explained in Sect. \ref{sec:contaminants}, the method mitigates for
this in a second pass.

This also illustrates that this method cannot be used to estimate
distances to individual sources because a given source with excess may
still have a solution from the model with an incorrect distance (although
the solution will be incorrect). A distance estimate can only be obtained
in the way we propose for equidistant and reasonably coeval
populations because it relies on the analysis of the
group. Consequently, the distance is more significantly
constrained for more populous or dense clusters.

In this sense, the effect of photometric incompleteness of the sample
is negligible. As long as the sample contains a large enough
number of coeval and equidistant sources to warrant sufficient
statistics relative to the background and foreground population, the
method is expected to be able to determine distances accurately. In the
current dataset we estimate that the cluster members represent 74\% of all
sources detected in $J$, $H,$ and $K_S$ simultaneously, even though the
photometric completeness is limited because of the $J$-band
sensitivity (see Sect. \ref{sec:final-sample}).

This method provides a good complement to {\it Gaia}
\citep{Gaia-Collaboration18} in determining distances to young
clusters. Because they are heavily reddened, these clusters often lack
sources visible in the optical ($\lambda < 1\mu$m), have only too
few of these sources to allow for a good handle on the distance to the
cluster, or are too distant so that the parallaxes need a higher-order
analysis to be converted into distances \citep{Luri18}. Our
method can be used reliably for embedded clusters and to reasonably
large distances.

\subsection{Distribution of excess}
\label{sec:distr-excess}

\begin{figure}
  \centering
  \resizebox{\hsize}{!}{\includegraphics{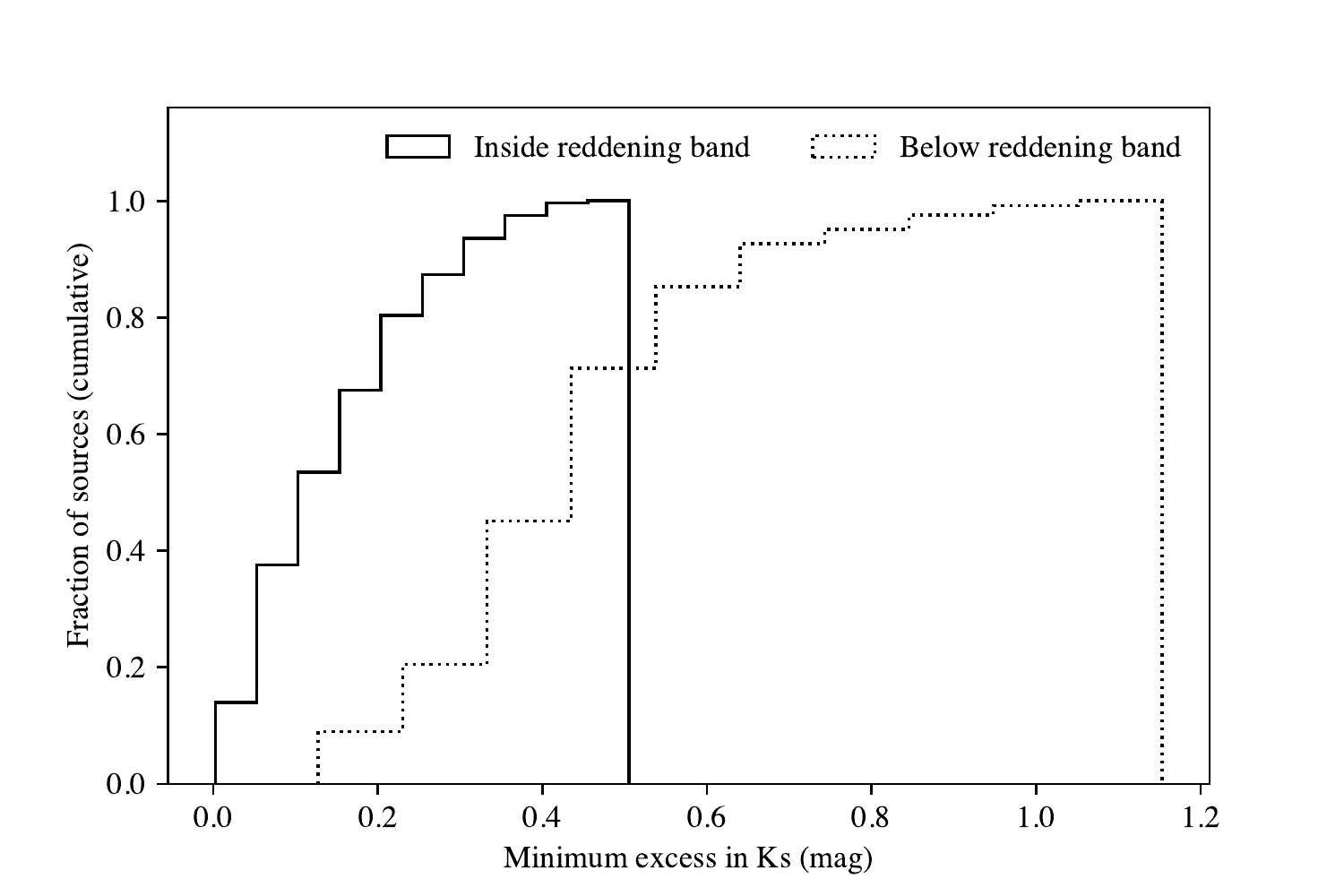}}
  \caption{Cumulative distribution of the minimum excess in $K_S$
    ($|e_K|$) required to find consistent colors and magnitudes for
    the sources whose colors lie inside the reddening band ({\it solid
      line}) and for those whose colors fall below the reddening band
    ({\it dotted line}).}
  \label{fig:ek_dist}
\end{figure}

The method we propose calculates the range of excess emission that is
required to produce consistent intrinsic colors and magnitudes for
each source. The actual amount of excess can be any value within these
ranges, and although the excess is not determined with our data, we
can characterize the distribution of the minimum excess required.

Figure \ref{fig:ek_dist} shows the cumulative distribution of the
minimum excess in $K_S$ required to fit sources inside the reddening
band (514 sources; {\it solid line}) and sources to the right of the
reddening band (122 sources; {\it dotted line}) for which the method
finds a solution. As expected, sources inside the reddening band need
lower excess on average than sources to the right of the reddening
band. In contrast to what is generally assumed, however, this figure
shows that sources inside the reddening band also need excess,
otherwise their photometry is irreconcilable with the models. For this
subset of sources, the figure shows that about 50\% need excess higher
than the maximum photometric error we allow for in our dataset.

Again, these distributions only show the lower limit of the excess
emission for these sources, and they therefore do not represent the
distribution of actual excess for this cluster, but they do show
that the excess distribution is probably not flat.

\subsection{Excess fraction of the cluster}
\label{sec:excess-fraction}

The lower limits for the amount of NIR excess for individual sources
(see Sect. \ref{sec:distr-excess}) can be used without further
assumptions to determine a lower limit for the excess fraction of the
cluster. In this way, this method becomes substantially more sensitive
than previous methods based on NIR data that select YSOs with excess
solely based on their binary position to the left or to the right of
the red limit of the reddening band in color-color space.

To account for the effect of photometric errors, we considered only
sources that require an excess higher than 0.15 magnitudes in $K_S$
($|e_{K,min}| \ge 0.15$).  Using this criterion, we find a lower limit
for the excess fraction in the $K_s$ band of 57\% (364 sources) for
RCW 38 in the science sample (see Sect. \ref{sec:contaminants}). In
comparison, only 19\% of the sources lie below the reddening band in
the \ccd. If we vary the distance between 1.3 and 2.0 kpc, and change
the age from 1 to 0.5 Myr, we obtain lower limits for the excess
fractions between 54\% and 59\%, confirming a very high excess
fraction for this cluster. With the extinction law coefficients from
\citet{Indebetouw05} or from \citet{Nishiyama09}, the excess fraction
becomes 59\% in both cases for an age of 1 Myr and a distance of 1.5
kpc.

Our estimate is a lower limit because the method returns a range of
possible excess for each source and we used the minimum value of
these ranges to define a source with excess. For example, if the
photometry of a given source can be reconciled with the models by
invoking an excess between 0.05 and 0.25 magnitudes, then this source
is not included as an excess source in the calculation of the excess
fraction, because the lower limit of the range is smaller than the
cutoff we defined of 0.15 magnitudes. However, if the true excess of
this source were closer to the upper limit of the range,
then this source would contribute to increase the excess fraction of
the cluster.

The incompleteness of the sample may affect the estimate of the excess
fraction of the cluster in the opposite direction. Excess increases
the brightness of YSOs by adding flux, which means that if some
low-mass YSOs that are not bright enough to be detected by themselves
will be detected if they have excess. This may bias the sample toward
sources with excess and artificially inflate the excess
fraction. However, this is a limited bias because it is not expected
that the emission from circumstellar material dominates the total
emission of the object in the NIR. Considering that excess is also
most likely red, that excess is expected to be prevalent in low-mass
YSOs, and that it is the $J$ band that limits the completeness the
most (see Fig.~\ref{fig:jhk_data}), we do not expect this effect to
bias the results significantly.

Keeping in mind that the minimum excess we refer to is not
necessarily the true excess of these sources (they are only lower
limits)  we searched for a dependence of minimum excess with position
within the cluster. This is a frail analysis with this dataset
because sensitivity and resolution vary spatially: as described in
Sect. \ref{sec:data}, the inner parts of the cluster were imaged with
higher resolution and sensitivity than the outer parts, and these
differences in completeness can influence the detection of
excess. With this in mind, we do not find that the sources that
require the highest excess avoid the proximity to the O-binary. This
may suggest that the massive stars have not (yet?) had a significant
impact on the $K$-band excess from the inner disk, or that dynamics
was already efficient in relocating stars from the place where their
disks were affected.

The excess fraction we derive is comparable to the fraction found
using MIR data with X-ray membership analysis for the same
cluster (70\%, although over a larger projected area and using
lower-resolution data; \citet{Winston11}). The gold standard for
detecting circumstellar material through excess emission is using
MIR observations, but this analysis suggests that this method
may be comparably sensitive, with the caveat that the near- and
MIR probe different regions and different properties of the
disk. This is a significant advance in the study of young clusters
because it opens the opportunity to analyze the presence of disks in
more challenging populations: current NIR instrumentation
delivers spatial resolutions that are several times better than in the
MIR. This allows studies of more distant clusters and of
clusters with higher stellar densities with an accuracy that is comparable to that of
studies of nearby, less massive and less dense environments, namely by
mitigating the effects of blended (unresolved) sources that can easily
inflate the number of excess sources artificially. Moreover, NIR
emission is still largely dominated by the object photosphere,
whereas the MIR is more sensitive to the emission from the
disks. NIR observations are therefore less likely to be
biased toward the detection of sources with disks, which would
artificially enhance excess fractions. Finally, NIR
observations are easily obtained using ground-based telescopes and
straightforward techniques, whereas MIR data require
space-based observations or expensive (in telescope time) ground-based
observations and sophisticated techniques to achieve comparable
performance because the sky is too bright in these wavelengths.

\subsection{Effect of unresolved binaries}
\label{sec:unresolved-binaries}

We have tested the method for the presence of unresolved binaries
using artificial star experiments. Any estimate of YSO parameters
(mass, excess and extinction) is incorrect for unresolved binaries if
the method provides a solution that is based on the expected
parameters for single sources, as the present method does.  However,
because it provides only lower and upper limits, the tests show that
these limits still encompass the mass and extinction of the binary.

The method was able to identify the vast majority of (cluster) YSOs
from background objects, both unresolved binaries and single sources,
in the artificial data. The method proved effective in finding the
correct distance in the artificial data, that is, in the presence of
unresolved binaries. The excess fraction changes when unresolved
binaries are included, although it remains underestimated. In other
words, the method still accurately predicts a reliable lower limit for
the fraction of excess sources in the cluster, even in the presence of
unresolved binaries, and even if the excess fraction is as high as
100\%.

The reason for this is that the combined magnitude and colors of two
blended sources with the same extinction (cluster members) is not too
different from those of each component, and the method can
nevertheless detect the excess. These conclusions hold as long as the
blending is not severe; if one detected source corresponds to many
YSOs due to a combination of poor resolution and large distance, then
more YSOs will be discarded from the sample, which may impact the
estimate of the excess fraction of the cluster, for example.

We therefore conclude that the method is robust against unresolved
binaries in determining global cluster parameters (distance and excess
fraction) and in detecting most cluster members if the blending is not
severe.

\subsection{Constraining masses}
\label{sec:constraining-masses}

This method provides constraints for the mass of YSOs,
as it does for excess, as we illustrated in Sect. \ref{sec:method}. The method returns an
interval of plausible masses for each source, and not a single most
likely value, as would be desirable.

Our results suggest that the upper limit of the mass interval
estimated for each source is likely an overestimate because it most
often corresponds to no excess in both $J$ and $H$ (and to the least
amount of excess in $K_S$). Although this is allowed by the conditions
we defined, most sources have a minimum excess in $K_S$ higher than
0.1 magnitudes, which means that the emitted flux
in $K_S$ of the circumstellar disk is more than 10\% of the flux emitted by the YSO itself. If a
YSO still has such a substantial disk, it is also likely to still have
a measurable excess (at least) in $H$ \citep[see
also][]{Furlan11}. This means that if we were to impose some
correlation between the excess in $K_S$ and in $H$, then the upper
limit for the mass of most sources would be lower. If we
adopt the slope of \citet{Meyer97} as the relation between color
excess in $(J-H)$ and $(H-K_S)$, then the upper limits for the mass
ranges allowed are lower than if we do not impose any relation between
the color of the excess emission, although we find that this relation is
not suitable for all objects in the sample.

The results provided by this method can unfortunately not be used to
reasonably constrain the mass function (MF) of the cluster because
the allowed mass intervals for each source are too wide to allow a
meaningful constraint. This method might at most be used to rule out
some extreme forms of the MF by sampling the allowed intervals for
individual sources a large enough number of times using Monte Carlo
methods and statistics. This was not attempted in this work.

\subsubsection{Consequences of neglecting excess emission}
\label{sec:comparison}

\begin{figure}
  \centering
  \resizebox{\hsize}{!}{\includegraphics{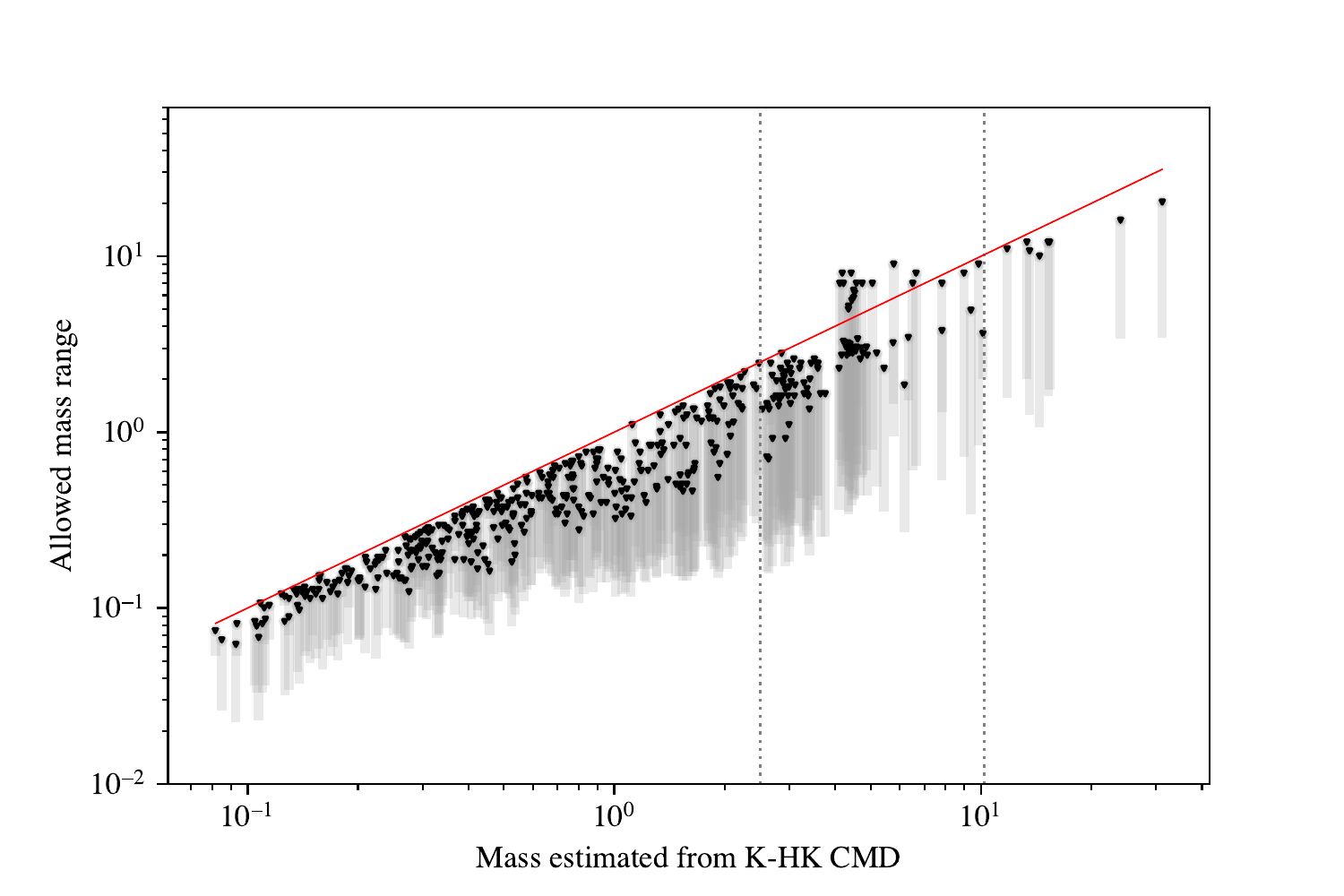}}
  \caption{Comparison between the masses estimated by the new method,
    allowing excess emission (allowed masses;
    Sect. \ref{sec:comparison}) and the masses inferred from the $K_S$
    \vs\ \hk\ CMD ($M_{\mathit{CMD}}$) assuming no excess
    emission. The vertical {\it gray bands} represent the range of
    allowed masses for each source, and the {\it inverted triangles}
    highlight the mass in this interval that corresponds to the
    least amount of excess emission. The horizontal error bar reflects
    the uncertainty in mass from photometric errors, except for the
    points inside the area delimited by the {\it dashed lines}, which
    corresponds to the range of masses in which the models are
    degenerate: in this area, the horizontal error bars encompass the
    minimum and maximum possible values from the models. The {\it 
      dashed red line} is the identity line.}
  \label{fig:comparison_masses}
\end{figure}

\begin{figure}
  \centering
  \resizebox{\hsize}{!}{\includegraphics{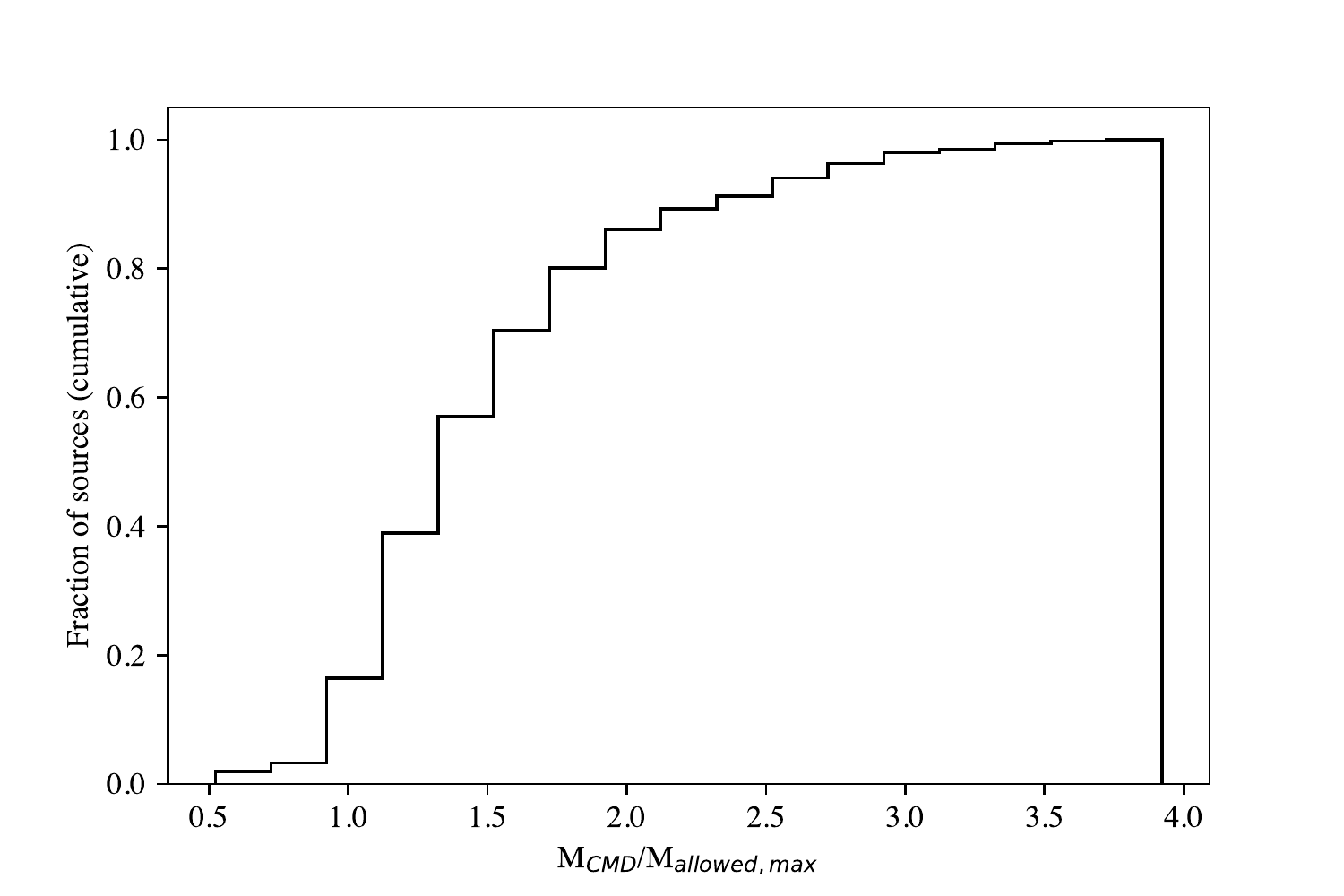}}
  \caption{Cumulative histogram of the ratio of the mass estimated
    directly from the CMD and the mass corresponding to the least
    excess for the sources inside the reddening band. This is a
    measure of the minimum factor by which the mass may be
    overestimated from the CMD.}
  \label{fig:comparison_masses_hist}
\end{figure}
 
Figure \ref{fig:comparison_masses} shows the comparison between the
masses estimated using the $K_S$ \vs\ \hk\ CMD assuming they do not
have excess emission and ignoring the inconsistency they present with
the colors in the CC diagram and those estimated using the method we
propose for the sources inside the reddening band. The vertical {\it
  gray bands} represent the range of allowed masses for each source,
that is, the masses corresponding to the green areas in
Fig. \ref{fig:method_redband}. As expected, compared to when we allow
sources to have excess, the CMD typically overestimates the mass, in
many cases by more than 100\% even for this subsample of sources
(sources inside the reddening band), because the excess emission is
taken as photospheric flux. The {\it inverted triangles} show the mass
that corresponds to the least amount of $K_S$-band excess emission,
which typically corresponds to the upper limit of the allowed mass
range for each source. Figure \ref{fig:comparison_masses_hist} shows a
cumulative histogram of the ratio of the mass estimated directly from
the CMD and the mass corresponding to the least excess ({\it inverted
  triangles} in Fig. \ref{fig:comparison_masses}); this ratio
corresponds to the minimum relative error in the mass of a source if
we take the mass from the CMD assuming no excess emission. The figure
shows, for example, that for 40\% of the sources, the CMD
overestimates the mass by at least 1.5 times relative to their allowed
masses.

\subsubsection{Using the $J$ \vs\ $(J-H)$ CMD to narrow down
  masses}
\label{sec:jjh_cmd}

The right panel of Fig. \ref{fig:method_redband} also shows the
mass and extinction estimates from a $J$ \vs\ $(J-H)$ CMD within the
photometric errors ({\it filled yellow squares}) for the example
source. Based on the typical spectral energy distribution of the
excess emission and our assumptions in particular, this band and this
color should be the least affected by excess emission. As
long as the excess is low, this color-magnitude space therefore is the most
likely to provide the correct estimate for the mass of each
source.  Fig. \ref{fig:method_redband} shows that the mass and extinction estimates from this
color-magnitude space for
one example source overlap the range of allowed masses
derived in Sect. \ref{sec:allowing-excesses} for almost all sources
(84\%). Unsurprisingly, they overlap the estimates that
correspond to the least amount of $K_S$-band excess emission. The question is whether this means that dereddening the sources to the models in the $J$ \vs\
$(J-H)$ CMD provides the correct mass, at least for the sources inside
the reddening band. This is not necessarily the case.

As we described in the previous section, the upper limit of the mass
estimate for each source is likely an overestimate of the true
mass. Furthermore, the estimate from the $J$ \vs\ $(J-H)$ CMD is still
inconsistent with the estimate from the \ccd\ ({\it yellow vs. blue}
points in the right panel Fig. \ref{fig:method_redband}). This can
only happen if there is excess exclusively in $K_S$, that is, if the
excess in $J$ and in $H$ is zero: in this case, the $(H-K_S)$ color in
the CC diagram would be offset from the reddened photospheric color,
but the $(J-H)$ color would be correct, as would the $J$-band
magnitude. Although this is plausible for sources with low excess in
$K_S$ because the excess in $H$ and in $J$ would then be even lower,
it becomes unlikely for sources with relatively high $K_S$-band
excess. In Sect. \ref{sec:distr-excess} we showed that a significant
fraction of sources indeed require such $K_S$-band excess, suggesting
that the excess at least in $H$ is also likely to be measurable, which
in turn suggests that the $J$-$(J-H)$ CMD cannot typically estimate
the mass correctly. Rather, this ubiquitous overlap of the $J$-$(J-H)$
mass estimate and the range of allowed masses is an expected
consequence of the constraints we assumed
(Sect. \ref{sec:define-param-space}) because we allow the possibility
of having excess only in $K_S$. We can at best say that the
$J$-$(J-H)$ CMD provides a still unlikely upper limit for the mass of
the sources that lie inside the reddening band.

\subsection{Applicability to other star-forming regions}
\label{sec:universality}

This analysis was conducted using data for one specific region, RCW
38. This method can be applied to any NIR dataset of YSOs
containing $J$-, $H$-, and $K$-band photometry, however.

The method we propose does not introduce any new principle; we suggest
that the photometric information from the three bands $J$, $H,$ and $K$
for a given source should be used consistently, along with the
theoretical models, instead of using preferentially one photometric
band and/or color. In this sense, the dataset for RCW 38 is only an
example of how this can be done and an example for the use of
NIR photometry to characterize clusters.

As mentioned before, this method only allows determining
distances to equidistant and reasonably coeval populations. Since it
relies on the statistical significance of the number of cluster
members in the dataset, the distance is more significantly
constrained the more populous the population.

The realization that the mass estimates from several color and
magnitude combinations are inconsistent with each other is independent
of the targeted region. RCW 38 is a very young cluster, therefore the
probability of its sources having disks is higher than for older
clusters, and these results may therefore be more evident for this
region than they would be for an older population. They showed that it
is not reasonable to assume \textup{a priori} that sources inside the
reddening band do not have excess, however, and that this assumption
may lead to large errors in the estimate of mass, for example. If the
sources do not have excess, then this method will show that the colors
and fluxes are consistent with each other without having to invoke the
presence of disks.

One limitation of the method we propose is the necessity of $J$-band
photometry because this is the band that is most affected by
interstellar extinction. For high extinction values (and large
distances), the lowest-mass YSOs cannot be detected in $J$ with
reasonable observing times. This problem will be mitigated with the
advent of 30-meter-class telescopes, whose collecting power will
be substantially larger, but this will always remain a limitation
of this method. Still, compared to the limitations inherent to other
methods for determining distances, memberships, and excess fractions, and
considering that the alternative of using only the more sensitive $H$
and/or $K$ bands may produce flawed results
(cf. Sect. \ref{sec:comparison}), we accept this limitation as
necessary.

Like all other methods, this method relies on data with
adequate spatial resolution. An inadequate combination of large
distance, high stellar density, and poor spatial resolution will cause
cluster members to be blended with other cluster members or with
unrelated sources. In both cases, the estimates for the individual
properties will be off, as with any other method that cannot
distinguish the two objects. As mentioned in
Sect. \ref{sec:unresolved-binaries}, the method is robust against
unresolved binaries, but it cannot accommodate many blended YSOs or
blended sources consisting of YSOs and unrelated objects with very
different extinctions. In this case, the combined photometry will be
too disparate for the method to be able to converge for that source,
and it will be discarded from the cluster sample. This will affect membership
determination, and also the estimates for distance and excess
fraction. This method is therefore not appropriate for observations of
clusters with inadequate resolutions relative to their distance and
surface density.

We also caution that the sensitivity of the method to detecting YSOs
(cluster members) likely decreases as a function of cluster age; the
photometric properties of older YSOs are progressively more similar to
those of main-sequence stars, so that the method will converge more easily
for unrelated objects if it expects the YSOs to have photometry that
is not too different from field stars.

We recommend these results to be taken into account at the observing
proposal stage for other regions, such that datasets used for the
study of young clusters include all three standard photometric broad
bands $J$, $H,$ and $K$ with adequate exposure times. This calls for
comparatively longer observations in $J$, and with adequate spatial
resolution.

\section{Conclusions}
\label{sec:conclusions}

Our study showed that NIR data can be very powerful in analyzing young
stellar clusters, more so than they are usually given credit for.  The
main conclusions of this paper are listed below.

\begin{enumerate}
\item We propose a new method that uses information from $J$, $H,$ and
  $K_S$ bands simultaneously to determine the distance to a cluster,
  assess the membership of individual sources, determine the fraction
  of sources that have excess emission from the hot inner rim of the
  circumstellar disks, and constrain the masses, NIR excess,
  and foreground extinction of individual YSOs independently of
  other observations.

\item The method returns a lower limit to the fraction of sources with
  excess in the cluster that is comparable to the fraction derived
  from existing MIR surveys, but at a higher
  resolution. This new opportunity can be a powerful tool for
  understanding the impact of massive stars and cluster environments on
  the evolution of circumstellar material across the Milky Way. For
  RCW 38, we find a lower limit for the excess fraction in the $K_s$
  band of $\approx$60\%.

\item We find that masses derived from NIR colors and magnitudes can
  be substantially overestimated using procedures that do not take
the information from the three photometric bands into account. This
  happens when sources whose colors fall inside the reddening band in
  a color-color diagram are (incorrectly) assumed to have no excess
  emission. The method we propose can constrain the range of possible
  mass and excess emission for single sources, both those whose colors
  lie inside and those that lie below the reddening band in a \ccd.

\item These results affect studies of individual sources and
  population studies, in particular studies of the initial mass
  function (IMF) in young and embedded clusters, which naturally rely
  on an accurate handle on individual masses. The bias is stronger at
  the low end of the mass spectrum, where most IMF research of
  embedded clusters is currently focused.
  
\item The method presented here bodes well for future studies of young
  clusters across the Milky Way. The upcoming generation of
  ground-based telescopes, such as the ELT, will have NIR
  instrumentation that is suited to constrain distances and
  memberships of deeply embedded or obscured young cluster across the
  entire Galactic plane.

\end{enumerate}

Although these results are drawn from the analysis of one single
cluster, the procedure is general and can be applied to any catalog of
YSOs with photometry in $J$, $H$, and $K$, provided the sources are
reasonably coeval and equidistant. The gain in using this procedure
with respect to estimating masses directly from a single-band
luminosity or from a single combination of color and magnitude without
accounting for excess is greater for younger populations because
circumstellar material tends to dissipate over time.

 \begin{acknowledgements}
   This work had the financial support of FCT, Portugal (grant
   ref. SFRH/BPD/101562/2014 and FCT contract number
   UIDB/00099/2020). A warm thanks to J. Alves for very useful
   discussions that brought clarity to this work. Thank you also to
   P. Garcia for suggestions, and to the referee, Prof. Sebastian
   Ramirez Alegria, whose thoughtful comments contributed to improve
   the paper. Based on observations collected at the European Southern
   Observatory under ESO programmes 70.C-0400(A), 70.C-0729(A), and
   074.C-0728(A).
 \end{acknowledgements}

  \bibliographystyle{aa} 
  \bibliography{/Users/jascenso/Work/Research/Documents/new_bib} 

\begin{appendix}

\section{Understanding the inconsistency between the CMD and the CC
  diagram}
\label{sec:understanding-discrepancy}

In this section, we explore the possible causes for the inconsistency
found in the parameters estimated from the CMD and the CC diagram
(Sect. \ref{sec:red-flag}) for the sources inside the reddening
band. We argue that this inconsistency is not likely related with
sample contamination, or with assuming an incorrect age or distance for
the cluster.

\subsection{Effect of sample contamination}
\label{sec:contamination}

One explanation for the inconsistency between the CC and CMD estimates
could be that the sample is dominated by unrelated sources: unlike
cluster members, field stars are distributed over a range of distances
and have a range of ages, so that estimating their mass from one single
stellar evolution isochrone at a fixed distance would result in a
similar inconsistency, at least qualitatively.

According to the calculations in
Sect. \ref{sec:understanding-discrepancy}, the colors and magnitudes of only 16\% of the sources in
our field of view are consistent with the
models, therefore only they might be cluster members if this were the
correct explanation. If we include the 94 sources that lie to the
right of the reddening band as cluster members as well because these
sources must have excess and therefore must be young, the sample would still
only contain 27\% cluster members. In this scenario, we
would have a contamination of 73\% in this field. This sample in
particular is very unlikely to be so severely contaminated by
unrelated sources: the field of view is very narrow, and it is
centered on a cluster that is rich and dense and that is deeply
embedded in a molecular cloud. The molecular cloud creates a wall of
extinction, especially in the $J$ band, which naturally filters many
background contaminants. For these reasons, we instead expect that
most of the observed sources are cluster members. It is accordingly not
reasonable to accept such a high contamination fraction. We therefore
rule out that the main cause for the observed inconsistency between
$M_{\mathit{CC}}$ and $M_{\mathit{CMD}}$ is contamination of the
sample from unrelated objects.

\subsection{Effect of distance}
\label{sec:effect-distance}

An incorrect distance might explain the discrepancy between the
estimates of mass and luminosity from the CMD and the CC
diagram. Altering the assumed distance does not change the estimate
from the CC diagram, but shifts the models in magnitude in the CMD,
producing different estimates from the CMD.

\begin{figure}
  \centering
  \resizebox{\hsize}{!}{\includegraphics{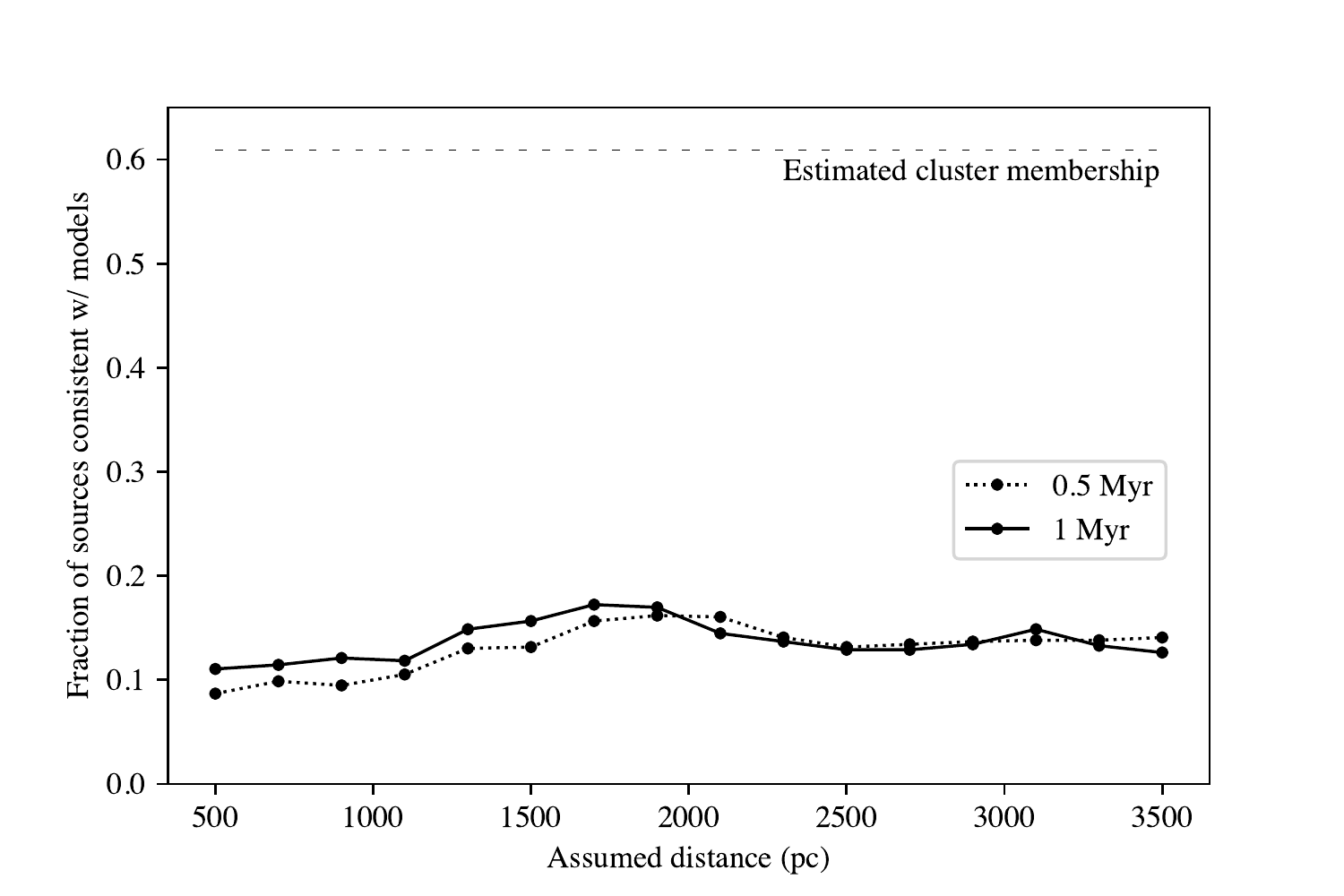}}
  \caption{Fraction of sources inside the reddening band for which the
    photometry can be reconciled with the models in color and
    magnitude within the photometric errors, assuming an age of 1.0
    Myr ({\it solid line}) or 0.5 Myr ({\it dotted line}). For
    reference, the {\it dashed line} shows the fraction of sources that we
    estimate are cluster members in Sect. \ref{sec:contaminants}.}
  \label{fig:check_distance}
\end{figure}

We investigated whether changing only the assumed distance (while
keeping the assumption that these sources do not have excess emission)
would increase the fraction of sources for which both colors and
magnitudes are consistent with the models and our current
assumptions. Fig. \ref{fig:check_distance} shows that this fraction
does not change significantly when the distance varies from 500 to 3500
pc: if an incorrect distance were the dominant cause for the
discrepancy, then this fraction would have a clear peak at the correct
distance because many sources (the cluster members) are expected to lie
at the same distance. Instead, the variations are within the expected
uncertainties from pure number statistics, and the fraction never
rises above 20\%. It is also implausible that these many YSOs are
spread over a wide range of distances because they are deeply
embedded in the molecular cloud. We therefore rule out that the observed
inconsistency between colors and magnitudes is due to an incorrect distance.

\subsection{Effect of age}
\label{sec:effect-age}

An incorrect age for the cluster might also cause a mismatch between
the CC and CMD mass estimates. There are several independent
indications that RCW 38 is a very young cluster (see
Sect. \ref{sec:introduction}), in particular younger than 1 Myr. We
therefore analyzed how changing the model isochrone (i.e., the
expected intrinsic colors and magnitudes as a function of mass) from 1
to 0.5 Myr would alter the fraction of sources for which the CC and
CMD mass estimates agree, while varying the distance as before.

We find that changing the assumed age changes the distribution for all
distances, but the values remain similar
(Fig. \ref{fig:check_distance}, {\it dashed line}). In particular,
there is also no distance for which more than 20\% of the sources agree between the CC and the CMD using the 0.5 Myr
isochrone. We therefore conclude that the observed inconsistency
between colors and magnitudes is not due to assuming an incorrect age for
the cluster either.

This confirmation was made assuming a single-age population. We therefore cannot
formally rule out the possibility that an age spread might explain the
inconsistency between the CC and CMD mass estimates. However, there is
no evidence in the literature that this cluster has a significant age
spread. Its deeply embedded state and its compact morphology,
added to the fact that we do not probe too far off the center of the
cluster with these data, argue against that for this dataset.

\end{appendix}

\end{document}